\documentclass[11pt]{article}
\usepackage{graphicx}
\usepackage[margin=1.25in]{geometry}
\usepackage[usenames,dvipsnames]{color}
\usepackage{url}
\usepackage[colorlinks = true,
            linkcolor = blue,
            urlcolor  = blue,
            citecolor = blue,
            anchorcolor = blue]{hyperref}
\usepackage{authblk}
\usepackage{arydshln}
\usepackage{amsmath,amssymb}
\usepackage{cite}
\usepackage{adjustbox}
\usepackage{multirow}

\def\ie{{\it i.e.}}
\def\eg{{\it e.g.}}

\def\beq{\begin{equation}}
\def\eeq#1{\label{#1}\end{equation}}
\def\eeqn{\end{equation}}

\newenvironment{Eqnarray}%
   {\arraycolsep 0.14em\begin{eqnarray}}{\end{eqnarray}}
\def\beqa{\begin{Eqnarray}}
\def\eeqa#1{\label{#1}\end{Eqnarray}}
\def\eeqan{\end{Eqnarray}}

\let\bar=\overbar

\def\lsim{\mathrel{\raise.3ex\hbox{$<$\kern-.75em\lower1ex\hbox{$\sim$}}}}
\def\gsim{\mathrel{\raise.3ex\hbox{$>$\kern-.75em\lower1ex\hbox{$\sim$}}}}

\def\O{{\cal O}}

\def\del{\partial}
\def\Dslash{\not{\hbox{\kern-4pt $D$}}}
\def\dslash{\not{\hbox{\kern-2pt $\del$}}}
\def\pslash{\not{\hbox{\kern-2pt $p$}}}
\def\ETmiss{\not{\hbox{\kern-4pt $E$}}_T}

\def\Dlr{\mathrel{\raise1.5ex\hbox{$\leftrightarrow$\kern-1em\lower1.5ex\hbox{$D$}}}}

\def\eff{{\mbox{\scriptsize eff}}}

\def\mz{m_Z}
\def\gz{\Gamma_Z}
\def\mw{m_W}
\def\gw{\Gamma_W}
\def\mt{m_t}

\def\mh{m_H}

\def\alphas{\alpha_s}
\def\MSB{{\bar{M \kern -2pt S}}}
\def\msb{{\bar{\scriptsize M \kern -1pt S}}}

\def\drb{{\bar{\scriptsize D \kern -1pt R}}}

\def\GeV{{\rm GeV}}
\def\TeV{{\rm TeV}}

\newcommand\snowmass{\begin{center}\rule[-0.2in]{\hsize}{0.01in}\\\rule{\hsize}{0.01in}\\
\vskip 0.1in Submitted to the  Proceedings of the US Community Study\\ 
on the Future of Particle Physics (Snowmass 2021)\\ 
\rule{\hsize}{0.01in}\\\rule[+0.2in]{\hsize}{0.01in} \end{center}}

\definecolor{darkred}{rgb}{0.5,0,0}

\begin{document}

\date{Version 2.4; October 24, 2022}

\title{{\normalsize\snowmass}
%EF04 Topical Group Report:
Electroweak Precision Physics and Constraining New Physics}

\author[1]{\textbf{Conveners:} Alberto Belloni\thanks{Alberto.Belloni@cern.ch}}
\author[2]{Ayres Freitas\thanks{afreitas@pitt.edu}}
\author[3]{Junping Tian\thanks{tian@icepp.s.u-tokyo.ac.jp}}

\author[4]{\newline\newline Juan Alcaraz Maestre}
\author[5]{Aram Apyan}
\author[6]{Bianca Azartash-Namin}
\author[7]{Paolo Azzurri}
\author[8]{Swagato Banerjee}
\author[9]{Jakob Beyer}
\author[10]{Saptaparna Bhattacharya}
\author[11]{Jorge de Blas}
\author[12]{Alain Blondel}
\author[13]{Daniel Britzger}
\author[14]{Mogens Dam}
\author[15]{Yong Du}
\author[16]{David d'Enterria}
\author[17]{Keisuke Fujii}
\author[9,18]{Christophe Grojean}
\author[19]{Jiayin Gu}
\author[2]{Tao Han}
\author[20]{Michael Hildreth}
\author[21]{Adri\'an Irles}
\author[16]{Patrick Janot}
\author[17]{Daniel Jeans}
\author[6]{Mayuri Kawale}
\author[22]{Elham E Khoda}
\author[23]{Samuel Lane}
\author[23]{Ian Lewis}
\author[24]{Zhijun Liang}
\author[9]{Jenny List}
\author[25]{Zhen Liu}
\author[2]{Yang Ma}
\author[26]{Victor Miralles}
\author[17,27]{Takahiro Mizuno}
\author[28]{Chilufya Mwewa}
\author[29]{Meenakshi Narain}
\author[30]{Luka Nedic}
\author[31]{Mark S.\ Neubauer}
\author[32,33]{Chaitanya Paranjape}
\author[34]{Michael Peskin}
\author[35]{Roberto Petti}
\author[28]{Marc-Andr\'e Pleier}
\author[36]{Roman Poeschl}
\author[30]{Karolos Potamianos}
\author[37]{Laura Reina}
\author[9]{J\"urgen Reuter}
\author[38]{Tania Robens}
\author[16]{Philipp Roloff}
\author[39]{Michael Roney}
\author[24]{Manqi Ruan}
\author[40]{Richard Ruiz}
\author[22]{Alex Schuy}
\author[41]{William Shepherd}
\author[33]{Daniel Stolarski}
\author[6]{John Stupak}
\author[42]{Taikan Suehara}
\author[7]{Roberto Tenchini}
\author[28]{Alessandro Tricoli}
\author[43]{Eleni Vryonidou}
\author[21]{Marcel Vos}
\author[6]{Connor Waits}
\author[23]{Graham Wilson}
\author[44]{Yongcheng Wu}
\author[2]{Keping Xie}
\author[45]{Siqi Yang}
\author[46]{Tianyi Yang}
\author[27]{Keita Yumino}
\author[47,48]{Shuang-Yong Zhou}
\author[49]{Junjie Zhu}

\affil[1]{University of Maryland, College Park, MD, USA}
\affil[2]{University of Pittsburgh, Pittsburgh, PA, USA}
\affil[3]{ICEPP, University of Tokyo, Japan}
\affil[4]{CIEMAT, Madrid, Spain}
\affil[5]{Brandeis University, Waltham, MA, USA}
\affil[6]{The University of Oklahoma, Norman, OK, USA}
\affil[7]{INFN Pisa, Italy}
\affil[8]{University of Louisville, Louisville, KY, USA}
\affil[9]{Deutsches Elektronen-Synchrotron DESY, 22607 Hamburg, Germany}
\affil[10]{Northwestern University, Evanston, IL, USA}
\affil[11]{CAFPE and Departamento de F\'isica Te\'orica y del Cosmos, Universidad de Granada, Spain}
\affil[12]{LPNHE, Paris, France}
\affil[13]{Max-Planck-Institut f\"ur Physik, Munich, Germany}
\affil[14]{Niels Bohr Institute, Copenhagen University, Denmark}
\affil[15]{Institute of Theoretical Physics, Chinese Academy of Sciences, Beijing, China}
\affil[16]{CERN, Geneva, Switzerland}
\affil[17]{IPNS, KEK, Tsukuba, Japan}
\affil[18]{Humboldt-Universit\"at zu Berlin, 12489 Berlin, Germany}
\affil[19]{Fudan University, Shanghai, China}
\affil[20]{University of Notre Dame, Notre Dame, IN, USA}
\affil[21]{IFIC (UV/CSIC) Valencia, 46980 Paterna, Spain}
\affil[22]{University of Washington, Seattle, WA, USA}
\affil[23]{University of Kansas, Lawrence, KS, USA}
\affil[24]{Institute of High Energy Physics, Chinese Academy of Sciences, Beijing, China}
\affil[25]{University of Minnesota, Minneapolis, MN, USA}
\affil[26]{INFN Rome, Italy}
\affil[27]{The Graduate University for Advanced Studies, SOKENDAI, Japan}
\affil[28]{Brookhaven National Laboratory, Upton, NY, USA}
\affil[29]{Brown University, Providence, RI, USA}
\affil[30]{Oxford University, Oxford OX1 3PJ, United Kingdom}
\affil[31]{University of Illinois at Urbana-Champaign, IL, USA}
\affil[32]{Indian Institute of Technology (Indian School of Mines) Dhanbad, Jharkhand-826004, India}
\affil[33]{Ottawa-Carleton Institute for Physics, Carleton University, Ottawa, ON, Canada}
\affil[34]{SLAC National Accelerator Lab., Menlo Park, CA, USA}
\affil[35]{University of South Carolina, Columbia, SC, USA}
\affil[36]{IJCLab,
CNRS / Universit\'e Paris-Saclay / Université Paris Cité, 91405 Orsay cedex, France}
\affil[37]{Florida State University, Tallahassee, FL, USA}
\affil[38]{Institut Ruder Boskovic, 10000 Zagreb, Croatia}
\affil[39]{University of Victoria, Victoria, BC, Canada}
\affil[40]{Institute of Nuclear Physics, Polish Academy of Sciences (IFJ PAN), Krak\'ow, Poland}
\affil[41]{Sam Houston State University, Huntsville TX, USA}
\affil[42]{Kyushu University, Fukuoka 819\u20130395, Japan}
\affil[43]{University of Manchester, Manchester M13 9PL, United Kingdom}
\affil[44]{Oklahoma State University, Stillwater, OK, USA}
\affil[45]{University of Science \& Technology, Hefei, China}
\affil[46]{Peking University, Beijing, China}
\affil[47]{ICTS, University of Science \& Technology, Hefei, China}
\affil[48]{Peng Huanwu Center for Fundamental Theory, Hefei, China}
\affil[49]{University of Michigan, Ann Arbor, MI, USA}

\medskip

\def\thefootnote{\fnsymbol{footnote}}
\setcounter{footnote}{0}

\maketitle

%\begin{abstract}
%\end{abstract}
\newpage
\tableofcontents

%%%%%%%%%%%%%%%%%%%%%%%%%%%%%%%%%%%%%%%%%%%%%%%%%%%%%%%%%%%%%%%%%%%%%%%%%%%
%%%%%%%%%%%%%%%% Introduction, big questions

\section{Introduction}

The precise measurement of physics observables and the test of their consistency within the standard model (SM) are an invaluable approach, complemented by direct searches for new physics, to determine the existence of physics beyond the standard model (BSM).
Historically, the discovery of new particles (\eg, the W and Z bosons by the UA1 and UA2 collaborations \cite{UA1:1983crd,UA2:1983tsx,UA1:1983mne,UA2:1983mlz}) has been followed by the construction of accelerator machines dedicated to the in-depth study of the new particles' features. After the discovery of a Higgs boson in 2012, there is no compelling theoretical argument or measurement result that predicts the mass scale of any BSM physics.
The indirect search for new physics, which exploits off-shell and loop contributions of new particles, allows one to explore a much wider range of energy scales than those probed by direct searches in specific BSM scenarios. Such indirect BSM effects are typically inversely proportional to some power of the mass scale of the new degrees of freedom \cite{Appelquist:1974tg}, so that high precision is crucial for probing large energy scales. The achievable precision of an experiment is determined by the statistics of the collected data sample, the experimental and theoretical systematic uncertainties, and their correlations. 

Studies of massive electroweak gauge bosons (W and Z bosons) are a promising target for indirect BSM searches, since the interactions of photons and gluons are strongly constrained by the unbroken gauge symmetries. They can be divided into two categories:

\begin{itemize}
\item Fermion scattering processes mediated by s- or t-channel W/Z bosons. These are known as \emph{electroweak precision measurements}, since large-statistics samples can be produced at $e^+e^-$ and $pp/p\bar{p}$ colliders. These measurements are sensitive to modifications of the gauge-boson--fermion couplings and the gauge-boson masses.

Electroweak precision tests at $e^+e^-$ colliders benefit from the clean and controlled initial state, whereas hadron colliders are affected by large systematic uncertainties due to parton distributions functions and other QCD effects. Thus $e^+e^-$ colliders have the potential to have a better sensitivity for electroweak precision measurements than hadron colliders, but a large integrated luminosity is crucial for that purpose.

Electroweak precision measurements will be covered in more detail in section~\ref{ewprec}, in particular the interplay of statistical, experimental systematic and theory uncertainties. It should be noted that it is very difficult to realistically predict the systematic uncertainties (both experimental and theory) of a future facility, since any uncertainty estimate is based on assumptions that can only be tested with data or by carrying out a certain theoretical calculation. Nevertheless, to fairly compare the potential of different proposed $e^+e^-$ colliders, the systematic uncertainties should be based on the same assumptions for all these machines. Such a consistent treatment of systematic error estimates has been attempted in this document.

\item Multi-boson processes, which include production of two or more vector bosons in fermion-antifermion annihilation, as well as vector boson scattering (VBS) processes. These processes can test modifications of gauge-boson self-interactions, and the sensitivity is typically improved with increased collision energy, so that hadron colliders tend to provide the strongest limits, although a future multi-TeV electron-positron or muon collider would also be very competitive.

A more extensive discussion of multi-boson physics at the high-luminosity run of the LHC (HL-LHC), at future higher-energy $pp$ colliders, and at high-energy $e^+e^-$ and $\mu^+\mu^-$ colliders is the topic of section~\ref{multbos}.
\end{itemize}

A model-independent description of indirect BSM effects is given by an extension of the SM with higher-dimensional operators. The most common effective theory framework for this purpose is the Standard Model Effective Field Theory (SMEFT), which has the same field content and symmetries as the SM. The leading contributions to electroweak observables stem from operators of dimension 6, which are suppressed by $\Lambda^{-2}$, where $\Lambda$ indicates an effective new physics scale. 

Generally, even at the dimension-6 level, there are more operators than independent observables, so that additional assumptions (\eg, about flavor symmetries) are needed to constrain the operator coefficient from the data. On the other hand, some of these operators also contribute to other phenomenological sectors of the SM, \ie, to Higgs physics or top physics, and measurements in these different sectors can help to break some parameter degeneracies. Thus it is advantageous to perform a \emph{global fit} of a large number of operators to a large number of observables from different sectors. In particular, such a global fit can be used to evaluate and compare the new physics reach of future experimental facilities.

Various global SMEFT fits of different scope are presented in section~\ref{glob}. Compared to previous such studies in the literature, the analysis in section~\ref{glob} uses updated inputs for the expected statistical and systematic uncertainties of key measurements at future colliders. Furthermore, it also extends previous studies by including 4-fermion operators, which generate contact interactions contributing to processes like $e^+e^- \to f\bar{f}$, and which can also modify the non-resonant background in Z-pole precision studies.

%%%%%%%%%%%%%%%%%%%%%%%%%%%%%%%%%%%%%%%%%%%%%%%%%%%%%%%%%%%%%%%%%%%%%%%%%%%
%%%%%%%%%%%%%%%% Electroweak precision

\section{Electroweak precision tests at future colliders}
\label{ewprec}

Precision measurements of the properties of W and Z bosons can be used to test the SM at the quantum level and to indirectly constrain potential BSM physics. The masses, widths and effective couplings of these gauge bosons can be modified through many different extensions of the SM, including new gauge interactions, extended Higgs sectors, composite Higgs scenarios, vector-like fermion fields, \emph{etc.} (\eg, see section 10 of Ref.~\cite{pdg2020} for an overview). 

\subsection{Current status of electroweak precision tests}
\label{sec:ewpocurr}

An important class of electroweak precision measurements focuses on fermion-pair production processes, $e^+e^- \to f\bar{f}$ and $pp \to \ell^+\ell^-$. For electron-positron colliders with center-of-mass energies near the Z-boson mass, the dominant contribution to the cross section follows from the Z resonance, which can be approximately written as
\begin{align}
    \frac{d\sigma}{d\Omega}[e^+e^-\to f\bar{f}] &\approx \frac{N_c^fs}{64\pi^2} \notag \\
     &\hspace{-5em} \times \frac{(1-P_+P_-)[G_1(1+c_\theta^2)+2G_3\,c_\theta]+(P_+-P_-)[H_1(1+c_\theta^2)+2H_3\,c_\theta]}{(s-\mz^2)^2+s^2 \gz^2/\mz^2},
\end{align}
where
\begin{align}
    G_1 &= (v_e^2+a_e^2)(v_f^2+a_f^2), &
    G_3 &= 4v_ea_ev_fa_f, \label{eq:Gpar} \\
    H_1 &= 2v_ea_e(v_f^2+a_f^2), &
    H_3 &= 2(v_e^2+a_e^2)v_fa_f, \label{eq:Hpar}
\end{align}
where $v_f$ and $a_f$ are the effective vector and axial-vector couplings of the Z-boson to the fermion type $f$, respectively, $N_c^f = 1\,(3)$ for leptons (quarks), and $P_{+/-}$ is the degree of longitudinal polarization of the positron/electron beam. Furthermore, $c_\theta \equiv \cos\theta$, where $\theta$ is the scattering angle.

The partial and total Z-boson widths can also be expressed in terms of these effective couplings,
\begin{align}
    \gz &= \sum_f \gz^f, & \gz^f &\approx \frac{N_c^f \mz}{12\pi}(v_f^2+a_f^2),
\end{align}
which in turn lead to the following expression for the total peak cross-section:
\begin{align}
    \sigma[e^+e^-\to f\bar{f}]_{s=\mz^2} &\approx \frac{12\pi}{\mz^2}\,\frac{\gz^e\gz^f}{\gz^2}.
\end{align}
The Z mass and total width can be determined from measurements of the cross-section lineshape at a few center-of-mass energies near the resonance peak. From measurements of cross sections with different final states one can determine ratios of the Z-boson partial widths. It is customary to express them in terms of
\begin{align}
    \sigma^0_{\rm had} &\equiv \sigma[e^+e^-\to\text{had.}]_{s=\mz^2}, &
    R_q &\equiv \frac{\Gamma_q}{\Gamma_\text{had}}\; (q=b,c), &
    R_\ell &\equiv \frac{\Gamma_\text{had}}{\Gamma_\ell} \; (\ell=e,\mu,\tau), \label{eq:sigR}
\end{align}
where ``had'' refers to all hadronic final states (\ie, the sum over $u,d,c,s,b$ final states at the partonic level).

Ratios of the vector and axial-vector couplings can be extracted from the forward-backward asymmetry, the average polarization degree of produced $\tau$ leptons (for $f = \tau$), and the left-right asymmetry (for a polarized electron beam): 
\begin{align}
    &\begin{aligned}
    A_{\rm FB} &\equiv \frac{\sigma_F-\sigma_B}{\sigma_F+\sigma_B} \approx \frac{3}{4}A_eA_f, \\
    \langle P_\tau \rangle &= A_\tau, \\
    A_{\rm LR} &\equiv \frac{\sigma_L-\sigma_R}{\sigma_L+\sigma_R} \approx A_e,
    \end{aligned}
    & A_f &\equiv \frac{2v_fa_f}{v_f^2+a_f^2}.
\end{align}
Here $\sigma_F$ and $\sigma_B$ refer to the cross section for only positive and negative values of $\cos\theta$, respectively; whereas $\sigma_L$ and $\sigma_R$ denote the cross section for left-handed ($P_-<0$) and right-handed ($P_->0$) electrons (assuming $P_+=0$). The ratio $v_f/a_f$ is also related to the \emph{effective weak mixing angle}
\begin{align}
    \sin^2\theta_\eff^f \equiv \frac{1}{4|Q_f|}\biggl(1-\frac{v_f}{a_f}\biggr).
\end{align}

The expressions above do not include the contributions stemming from photon-exchange and box diagrams and from radiative corrections that cannot be absorbed into the effective couplings, in particular initial-state and final-state radiation. These effects need to be predicted from theory and subtracted from the data in order to extract ``measured'' values of the quantities in eqs.~\eqref{eq:sigR}. The latter are therefore known as electroweak \emph{pseudo-observables} (EWPOs).

Other EWPOs include the W-boson mass ($\mw$) and branching ratios, as well as the Fermi constant of muon decay, $G_F$. The latter is a key ingredient for predicting $\mw$ in the SM, based on the relation
\begin{equation}
    \frac{G_F}{\sqrt{2}} = \frac{\pi\alpha}{2\mw^2(1-\mw^2/\mz^2)}(1+\Delta r),
\end{equation}
where $\Delta r$ describes higher-order corrections. $G_F$ is currently known with a precision of 0.5 ppm \cite{pdg2020}, which may be further improved in the future, and thus it is a negligible source of uncertainty.

Moreover, when comparing experimental values for the EWPOs to theory predictions in the SM, other SM parameters are needed as inputs for the latter. Specifically, the mass of the top quark and the Higgs boson play an important role, as well as the strong coupling constant $\alphas$ and the shift due to the running of the fine structure constant from the Thomson limit to the Z scale, 
$
    \Delta\alpha \equiv 1-\frac{\alpha(0)}{\alpha(\mz)}.
$
\label{deltaalpha1}
$\Delta\alpha$ receives contributions from leptons, which can be computed perturbatively \cite{Sturm:2013uka}, and from hadronic states. The hadronic part can be split into non-perturbative and perturbative contributions. The non-perturbative contributions can be extracted from data for $e^+e^- \to \text{had.}$ \cite{Blondel:2019vdq,Davier:2019can,Keshavarzi:2019abf} or from lattice QCD simulations \cite{Burger:2015lqa,Ce:2022eix} using a dispersive approach. The data-driven methods are currently more precise, with an uncertainty for $\Delta\alpha_{\rm had}$ of about $10^{-4}$ \cite{Blondel:2019vdq,Davier:2019can,Keshavarzi:2019abf}.

Reducing the uncertainty of $\Delta\alpha_{\rm had}$ requires improved measurements of $e^+e^- \to \text{had.}$ for energies below 2 GeV (\eg, with ongoing measurements at VEPP-2000 and BEPC-II), 4-loop perturbative QCD corrections, and more precise determinations of the charm and bottom quark masses. With these improvements, an uncertainty of $<0.5 \times 10^{-4}$ appears within reach \cite{Blondel:2019vdq}. Similarly, the lattice QCD evaluation of $\Delta\alpha_{\rm had}$ is expected to continue to improve, but quantitative estimates are currently not available.

%-------------------------------------------------------------------------
\begin{table}[tb]
  \begin{center}
  \begin{tabular}{|c|c|c|}
    \hline
    EWPO Uncertainties        &    Current      & HL-LHC     \\
    \hline
    $\Delta\mw$ (MeV)          &    12 / 9.4$^\dag$  &    5      \\
    $\Delta\mz$ (MeV)          &    2.1  &     \\
    $\Delta\gz$ (MeV)     &    2.3    &  \\ 
    $\Delta\mt$ (GeV)    &  0.6* &  0.2 \\
    \hdashline
    $\Delta\sin^2\theta^\ell_\eff$ ($\times 10^5$)   &    13 & $<10$     \\ 
    $\delta R_\mu\;(\times10^3)$  &    1.6    &      \\
    $\delta R_b\;(\times10^3)$    &    3.1    &      \\
    \hline    
  \end{tabular}\\[2ex]
  $^\dag$ \parbox[t]{.6\textwidth}{\footnotesize{The recent W mass measurement from CDF with 9.4~MeV precision \cite{CDF:2022hxs} has not yet been included in the global average \cite{pdg2020}.}}\\[1ex]
  * \parbox[t]{.6\textwidth}{\footnotesize{This value includes an additional uncertainty due to ambiguities in the top mass definition (see EF TOPHF report \cite{Schwienhorst:2022hht} for more details).}}
  \end{center}
\vspace{-2ex}
\caption{The current precision of a few selected EWPOs, based on data from LEP, SLC, TeVatron and LHC \cite{pdg2020}, and expected improvements from the HL-LHC \cite{HLLHC}. $\Delta$ ($\delta$) stands for absolute (relative) uncertainty.}
\label{tab:ewpocurr}
\end{table}
%-------------------------------------------------------------------------
The current precision for a few selected EWPOs is listed in Tab.~\ref{tab:ewpocurr}. Most of these results stem from measurements at the $e^+e^-$ colliders LEP and SLC, but the hadron colliders TeVatron and LHC contribute important results for $\sin^2\theta^\ell_\eff$, $\mw$ and $\mt$.

The HL-LHC with integrated luminosity of 3000~fb$^{-1}$ can make improved measurements of certain EWPOs. The effective weak mixing angle can be extracted from measurements of the forward-backward asymmetry in Drell-Yan production, $pp \to \ell^+\ell^-$ ($\ell=e,\mu$). The measurement precision is mostly limited by uncertainties of the parton distribution functions (PDFs), but the PDFs can be constrained simultaneously with the weak mixing angle determination through Drell-Yan data. In particular, the $m_{\ell\ell}$ distribution can be useful in disentangling the effect of PDFs from the weak mixing angle determination \cite{Yang:2021cpd}.
It is estimated that the total uncertainty of $\sin^2\theta^\ell_\eff$ can be reduced below $10^{-4}$ at HL-LHC \cite{HLLHC}.

Similarly, the W-boson mass can be extracted from measurements of $pp \to \ell\nu$, by performing fits to the lepton $p_T$ and transverse mass distributions. This measurement benefits from a dedicated run with low instantaneous luminosity to improve the accurate reconstruction of the missing transverse momentum. Again PDF uncertainties are expected to dominate, and an ultimate precision of about 5 MeV appears achievable \cite{HLLHC}. 

It should be noted that these precision measurements at the HL-LHC will rely on detailed theory input, including higher-order EW and mixed QCD$\times$EW corrections \cite{Buonocore:2021rxx,Bonciani:2021zzf,Buccioni:2022kgy}, as well as resummation for low $p_T$ (\eg, see Ref.~\cite{Camarda:2021ict} and references therein).
Moreover, the extraction of $\sin^2\theta^\ell_\eff$ assumes that the dependence of $\sin^2\theta^f_\eff$ on different fermion flavor $f$ is small and as predicted in the SM\footnote{In other words, this measurement can serve as a high-precision consistency check of the SM, but it will be difficult to interpret an observed deviation from the SM without model assumptions.}.

\subsection{Electroweak precision measurements at future \boldmath $e^+e^-$ colliders}

Future high-luminosity $e^+e^-$ colliders proposed as Higgs factories can also be used to study the masses and interactions of electroweak bosons to much higher precision than before. We here focus on four collider proposals: ILC \cite{Baer:2013cma,Bambade:2019fyw,ILCInternationalDevelopmentTeam:2022izu}, CLIC \cite{Linssen:2012hp,Charles:2018vfv}, FCC-ee \cite{Abada:2019zxq,Bernardi:2022hny}, and CEPC \cite{CEPCStudyGroup:2018ghi,Gao:2022lew,Cheng:2022zyy}. 
Table~\ref{tab:eecoll} summarizes the run scenarios considered for these colliders within the Snowmass 2021 study. The 50~MW upgrade of CEPC \cite{Cheng:2022zyy} is assumed for all quantitative analyses throughout this document. The recent Cool Copper Collider (C$^3$) \cite{Bai:2021rdg,Dasu:2022nux} proposal has parameters very similar to ILC and will not be discussed separately in what follows.
%-------------------------------------------------------------------------
\begin{table}[tb]
\begin{center}
\begin{tabular}{|l|c|c|c|}
\hline
Collider & $\sqrt{s}$ & P [\%] & $L_{\rm int}$ \\
         &            & $e^-/e^+$ & ab$^{-1}$ \\[.5ex]
\hline
ILC & 250 GeV & $\pm 80 / \pm 30$ & 2 \\
    & 350 GeV & $\pm 80 / \pm 30$ & 0.2 \\
    & 500 GeV & $\pm 80 / \pm 30$ & 4 \\
    & 1 TeV   & $\pm 80 / \pm 20$ & 8 \\[1ex]
\hline
ILC-GigaZ & $\mz$ & $\pm 80 / \pm 30$ & 0.1 \\[1ex]
\hline
CLIC& 380 GeV & $\pm 80 / 0$ & 1 \\
    & 500 GeV & $\pm 80 / 0$ & 2.5 \\
    & 1 TeV   & $\pm 80 / 0$ & 5 \\[1ex]
\hline
CEPC& $\mz$ & & 60 / 100 \\
    & $2\mw$ & & 3.6 / 6 \\
    & 240 GeV & & 12 / 20 \\
    & $2\mt$ & & -- / 1 \\[1ex]
\hline
FCC-ee& $\mz$ & & 150 \\
    & $2\mw$ & & 10 \\
    & 240 GeV & & 5 \\
    & $2\mt$ & & 1.5 \\[1ex]
\hline
\end{tabular}
\end{center}
\vspace{-2ex}
\caption{Electron-positron collider run scenarios used for the Snowmass 2021 study. The two sets of numbers for CEPC refer to the 30 MW baseline and 50 MW upgrade for the beam power. Also see section 1.3 and table 1-1 in the main EF report \cite{Narain:2022qud}.}
\label{tab:eecoll}
\end{table}
%-------------------------------------------------------------------------

The linear collider projects ILC and CLIC feature polarized electron beams (and also a polarized positron beam in the case of ILC).
Two options are considered for ILC, the default option with center-of-mass energies of 250 GeV and above, and the ``GigaZ'' option that includes a run at the Z pole. [Note that a Z-pole run is also considered as a possible option for CLIC \cite{Gohil:2687090}.] The ILC and CLIC runs with 500 GeV and above are irrelevant for ``canonical'' electroweak precision studies (\ie, not considering multi-gauge-boson processes).

The circular colliders (FCC-ee and CEPC) can deliver very large integrated luminosities on the Z pole, yielding samples of $\O(10^{12})$ events. On the other hand, the standard run scenarios for ILC (without the GigaZ option) and CLIC do not include any run at the Z pole. Instead, precision studies of the Z boson are possible through the radiative return method, \ie, by producing Z bosons together with one or more initial-state photons, $e^+e^- \to Z+n\gamma$. The photons are emitted predominantly at low angles and lost in the beam pipe. However, the requirement of hard photon emission reduces the event yield and thus the achievable precision.

Given the large statistics of these future $e^+e^-$ colliders, systematic uncertainties may have a significant impact on the achievable precision. In the following we discuss the dominant machine-specific systematic error sources, as well as uncertainties that are common to all machines. It should be emphasized that these systematic error evaluations are just order-of-magnitude estimates, while a more precise assessment would require instrumentation detail and tools that are not available at this time.

\paragraph{Common systematics:} Uncertainties due to the physics modeling affect all collider proposals equally. Previous publications by the collider collaborations \cite{LCCPhysicsWorkingGroup:2019fvj,Blondel:2021ema,CEPCStudyGroup:2018ghi,RoloffEF04} have made differing assumptions for the size and relevance of these common uncertainties. This situation creates problems for the comparison of the new-physics reach between different machines. Therefore, as part of the Snowmass 2021 process, a consistent set of assumptions is being used and applied uniformly to all $e^+e^-$ collider proposals.

For branching ratios of heavy-quark (b and c) final states, the tagging efficiency can be controlled \emph{in situ} by comparing single and double tag rates. However, the simple scaling $\epsilon_{\rm 2tag} = (\epsilon_{\rm 1tag})^2$ gets modified by so-called hemisphere correlations. These correlations can be produced by detector effects, vertex fitting, and QCD effects. The first two sources can be reduced to a sub-dominant level through the availability of large-statistics calibration samples and the increased vertex precision of modern vertex detectors. The most important QCD effect is gluon splitting into a heavy-flavor $q\bar{q}$ pair. The contamination from gluon splitting can be reduced with acolinearity cuts between the two tagged jets. Moreover, the large available data sets can be used to dramatically improve the modeling of gluon splitting, but this will require parallel improvements in Monte-Carlo (MC) simulation tools. Here it is assumed that the QCD uncertainty can be improved by about one order magnitude compared to LEP \cite{ALEPH:2005ab}, leading to relative uncertainties of $0.2 \times 10^{-3}$ for $R_b$ and $1\times 10^{-3}$ for $R_c$, respectively.

Similarly, QCD effects are a dominant source of uncertainty for determinations of $A_b$ ($A_c$) from the forward-backward asymmetry of $e^+e^- \to b\bar{b}$ ($c\bar{c}$). QCD radiation can change the angular distributions and correlations of the heavy-quark jets, which in turn modified the observable asymmetry. This has been studied in detail in Ref.~\cite{AlcarazMaestre:2020fmp}, where it was found that the impact of QCD effects can be substantially reduced with an acolinearity cut. With a moderate acolinearity cut and assuming NNLO QCD corrections, the relative error on $A_{b,c}$ due to missing higher-order perturbative QCD contributions is estimated to be about $3 \times 10^{-4}$ (see Tab.~9 in Ref.~\cite{AlcarazMaestre:2020fmp}). With future work on QCD calculations this can likely be reduced to the level of $1\times 10^{-4}$. In addition, one needs to consider non-perturbative hadronization and showering uncertainties (see also Ref.~\cite{dEnterria:2020cgt}). Due to wealth of available data at any of the proposed colliders, a significant improvement of the hadronization and showering models should be possible. Assuming an improvement of a factor 5 compared to currently available MC tunes (see Tab.~9 in Ref.~\cite{AlcarazMaestre:2020fmp}), this leads to an estimated relative error of $2\times 10^{-4}$. Combining perturbative and non-perturbative uncertainties, the total absolute error due to QCD effects amounts to $2.1 \times 10^{-4}$ for $A_b$ and $1.5 \times 10^{-4}$ for $A_c$.

\paragraph{Experimental systematic for linear colliders:} For electroweak precision measurements at ILC250 or CLIC380 using the radiative return method, signal events need to be selected based on the invariant mass $m_{ff}$ of two fermions from $Z \to f\bar{f}$. $m_{ff}$ can be reconstructed using the polar angles of the fermions \cite{ALEPH:1998ac}, which can be measured very precisely, so that this becomes a negligible source of systematic uncertainty. Note that multi-photon emission produces a tail in the reconstructed $m_{ff}$ distribution, but this dilution does not diminish the precision of the overall energy scale calibration.

Combining this method with a precise calibration of the tracker momentum scale using large samples of kaon and $\Lambda$ baryon decays, it may be possible to determine the Z mass and width at ILC250 with a systematic uncertainty of 2 ppm \cite{WilsonEF04}.

Many measurements at linear $e^+e^-$ colliders profit from polarized beams, which in turn makes the polarization calibration a leading source of systematic uncertainties. Both the ILC and CLIC designs expect that the luminosity-weighted long-term average of the polarization can be controlled to 0.1\% \cite{ILCInternationalDevelopmentTeam:2022izu,Linssen:2012hp}. However, as demonstrated in Ref.~\cite{Beyer:2022nyr}, the impact of the polarization uncertainty can be further reduced by treating the polarization values as nuisance parameters in the actual extraction of physics parameters from a set of observables.

The asymmetry parameter $A_e$ can be determined from the left-right asymmetry $A_{\rm LR}$ (see Ref.~\cite{Mizuno:2022xuk} for a full simulation study), while $A_f$ for other fermion types ($f=\mu,\tau,b,c$) can be obtained from the left-right-forward-backward asymmetry for the process $e^+e^- \to f\bar{f}$,
\begin{align}
A_{\rm LR,FB} = \frac{\sigma_{LF} - \sigma_{LB} - \sigma_{RF} + \sigma_{RB}}{\sigma_{LF} + \sigma_{LB} + \sigma_{RF} + \sigma_{RB}} \approx \frac{3}{4}A_f
\end{align}
The polarization calibration leads to a relative systematic uncertainty of $3\times 10^{-4}$. Other important systematic uncertainties include the control of the luminosity and detector acceptance between runs with different polarization, which are estimated to be subdominant.

For the branching ratios $R_i$, the dominant source of uncertainty stems from the flavor identification, which is estimated at the level of 0.1\% \cite{ILCInternationalDevelopmentTeam:2022izu}.

Measurements of $e^+e^- \to W^+W^-$ at ILC250 or CLIC380 can yield information about a variety of properties of the W bosons, including anomalous gauge-boson couplings. The W mass can be determined from a variety of kinematic final-state observables \cite{ILCInternationalDevelopmentTeam:2022izu}: (1) constrained reconstruction of $qq\ell\nu$ events; (2) di-jet invariant mass for semi-leptonic and all-hadronic final states; (3) endpoints of the lepton energy spectrum for di-lepton ($\ell\nu\ell\nu$) and semi-leptonic ($qq\ell\nu$) final states ; and (4) approximate kinematic reconstruction of di-lepton events by assuming that the event has a planar topology (``pseudo-mass'' method). With an integrated luminosity of a few ab$^{-1}$ at ILC250, a statistical uncertainty of 0.5~MeV on $\mw$ can be achieved \cite{ILCInternationalDevelopmentTeam:2022izu}. The systematic uncertainty was estimated in the Snowmass 2013 study with 2.4~MeV \cite{Baak:2013fwa}, which receives comparable contributions from the beam-energy calibration, luminosity spectrum, modeling of hadronization, modeling of radiative corrections, and detector energy calibration. With improved detectors and methods for addressing the other systematic issues, a total error of 1~MeV at ILC250 may well be feasible.

An ILC run on the Z pole (ILC-GigaZ) would yield a higher-statistics sample of clean Z events, thus leading to improved overall precision for EWPOs. For the asymmetry parameters $A_f$, the systematic errors are again dominated by the polarization uncertainty, whereas the acceptance is the leading source of systematics for the branching ratios $R_f$ \cite{LCCPhysicsWorkingGroup:2019fvj}.

\paragraph{Experimental systematic for circular colliders:} The beam energy at circular colliders can be controlled with high accuracy using resonant depolarization, leading to an absolute precision of 100 keV (for $\sqrt{s} \sim 100$~GeV) and a point-to-point precision of 25~keV (\ie\ the accuracy with which the energy difference between two nearby center-of-mass energies can be determined). These two numbers are the leading systematic uncertainties for the determination of the Z mass and width, respectively.

For the determination of $A_e$, it is advantageous to consider the forward-backward tau polarization in $e^+e^- \to \tau^+\tau^-$,
\begin{align}
\langle P_{\rm \tau,FB} \rangle &\equiv \frac{\langle P_\tau\rangle_F-\langle P_\tau\rangle_B}{\langle P_\tau\rangle_F+\langle P_\tau\rangle_B} \approx \frac{3}{4}A_e\,.
\end{align}
This quantity is independent of the tau polarization distributions and of hemisphere migration effects. It would only be affected by correlations between these two effects, which are expected to be very small. As a result, the dominant systematic uncertainty would instead stem from non-tau backgrounds. These are estimated from the statistics of control samples used for calibrating the background, leading to an error estimate of $2\times 10^{-5}$.

Other $A_f$ parameters can be determined from $A_{\rm FB}$ for $e^+e^- \to f\bar{f}$. The main systematic uncertainty for $A_\mu$ is point-to-point control of the luminosity and detector acceptance. For $A_c$ and $A_b$ the dominant systematic error stems from QCD effects (see ``Common systematics'' above). While $A_\tau$ could also be obtained from the forward-backward asymmetry, a more precise determination is possible from the tau polarization. The main systematic uncertainty for the polarization measurement is due to the modeling of the hadronic tau decay modes. Since these are expected to be substantially improved by using the large available calibration samples at FCC-ee/CEPC, it is expected that this error is reduced by a factor 10 compared to LEP \cite{ALEPH:2005ab}, leading to uncertainty of $2 \times 10^{-4}$.
%%% Note: we did not get sufficient info from CEPC for the A_\tau systematics, and thus we took the FCC-ee value

The measurement of the total peak cross-section, $\sigma^0_{\rm had}$, is limited by the luminosity calibration. Using low-angle Bhabha events, a relatively precision of $10^{-4}$ or better should be achievable. For the hadronic branching fractions $R_{b,c}$, QCD uncertainties from gluon splitting are the dominant error source (see ``Common systematics'' above). For the leptonic branching fractios $R_{e,\mu,\tau}$ the lepton acceptance and beam energy control will be important factors. $R_e$ is additionally affected by the subtraction of Bhabha backgrounds.

The W mass and width can be extracted with high precision from measurements at a few energy points near the $WW$ threshold. As mentioned above, the dominant systematic uncertainty due to the beam energy calibration can be controlled using resonant depolarization, but with a slightly lower precision (0.0002\%) at this center-of-mass energy compared to the Z pole.

%-------------------------------------------------------------------------
\begin{table}[tb]
  \begin{adjustbox}{max width = \textwidth}
  \begin{tabular}{|c|c|c|c|c|c|c|}
    \hline
    Quantity       & current  &    ILC250      & ILC-GigaZ   &      FCC-ee            &  CEPC      &  CLIC380     \\
    \hline
    $\Delta\alpha(\mz)^{-1}\;(\times 10^3)$ &    17.8$^*$  & 17.8$^*$  &             &    3.8 (1.2)  & 17.8$^*$ &           \\
    $\Delta\mw$ (MeV)     & 12$^*$     &    0.5 (2.4)  &             &    0.25 (0.3)   & 0.35 (0.3)   &           \\
    $\Delta\mz$ (MeV)    & 2.1$^*$      &    0.7 (0.2)  &  0.2           &    0.004 (0.1)   & 0.005 (0.1)   &    2.1$^*$       \\
    $\Delta\mh$ (MeV)   & 170$^*$       &    14  &             &    2.5 (2)   & 5.9   &    78       \\
    $\Delta\gw$ (MeV)     & 42$^*$ &    2   &             &    1.2 (0.3)   & 1.8 (0.9)   &           \\    
    $\Delta\gz$ (MeV)     & 2.3$^*$ &    1.5 (0.2)  &  0.12    &    0.004 (0.025)   & 0.005 (0.025)   &    2.3$^*$       \\ 
    \hdashline
    $\Delta A_e\;(\times 10^5)$   & 190$^*$ &    14 (4.5)  &   1.5 (8)  &    0.7 (2)   & 1.5 (2)   &    60 (15)       \\ 
%%% Note: we did not get sufficient info from CEPC for the A_e systematics, and thus we took the FCC-ee value
    $\Delta A_\mu\;(\times10^5)$  & 1500$^*$ &    82 (4.5)  &   3 (8)  &    2.3 (2.2)   & 3.0 (1.8)   &    390 (14)      \\    
    $\Delta A_\tau\;(\times10^5)$ & 400$^*$ &   86 (4.5)  &   3 (8)  &    0.5 (20)   & 1.2 (20)   &    550 (14)      \\        
%%% Note: we did not get sufficient info from CEPC for the A_e systematics, and thus we took the FCC-ee value
    $\Delta A_b\;(\times10^5)$   & 2000$^*$ &   53 (35)  &   9 (50)  &    2.4 (21)   & 3 (21)  &    360     (92)  \\
    $\Delta A_c\;(\times10^5)$  & 2700$^*$  &  140 (25)  &  20 (37)  &    20 (15)   & 6 (30)   &    190     (67)  \\ 
%%% Note: we would still like to understand by CEPC has a smaller statistical error but a larger systematic error than FCC-ee
    \hdashline
    $\Delta \sigma_{\rm had}^0$ (pb)  & 37$^*$ &           &          &    0.035 (4)      & 0.05 (2)     &    37$^{*}$       \\     
    $\delta R_e\;(\times10^3)$    & 2.4$^*$ &    0.5 (1.0)    &   0.2 (0.5)  &    0.004 (0.3)   & 0.003 (0.2)   &    2.5 (1.0)       \\
    $\delta R_\mu\;(\times10^3)$  &  1.6$^*$  & 0.5 (1.0)    &   0.2 (0.2)  &    0.003 (0.05)  & 0.003 (0.1)   &    2.5 (1.0)       \\
    $\delta R_\tau\;(\times10^3)$ &  2.2$^*$ &  0.6 (1.0)    &   0.2 (0.4)  &    0.003 (0.1)   & 0.003 (0.1)   &    3.3 (5.0)       \\
    $\delta R_b\;(\times10^3)$    & 3.1$^*$ &   0.4 (1.0)    &   0.04 (0.7)  &   0.0014 ($<0.3$)   & 0.005 (0.2)   &    1.5 (1.0)       \\
    $\delta R_c(\times10^3)$    &  17$^*$ &  0.6 (5.0)    &   0.2 (3.0)  &    0.015 (1.5)   & 0.02 (1)   &    2.4 (5.0)       \\
    \hline    
  \end{tabular}
  \end{adjustbox}
\vspace{-2ex}
\caption{EWPOs at future $e^+e^-$ colliders: statistical error (estimated experimental systematic error). $\Delta$ ($\delta$) stands for absolute (relative) uncertainty, while * indicates inputs taken from current data \cite{pdg2020}. See Refs.~\cite{deBlas:2019rxi,DeBlas:2019qco,LCCPhysicsWorkingGroup:2019fvj,Blondel:2021ema,ILCInternationalDevelopmentTeam:2022izu,Cheng:2022zyy}.}
\label{tab:ewpocomp}
\end{table}
%-------------------------------------------------------------------------

\paragraph{Summary:} A summary of projected statistical and systematic uncertainties for the different proposed $e^+e^-$ colliders is given in Tab.~\ref{tab:ewpocomp}. This table also serves as an input for the global fits presented in section~\ref{glob}.

Measurements of leptonic branching ratios $R_\ell$ ($\ell = e,\mu,\tau$) can be used to extract a precise value for the strong coupling constant $\alphas$, which enters through final-state radiative corrections in $\Gamma_{\rm had}$. Given that $R_\ell$ is a highly inclusive quantity, this determination of $\alphas$ is essentially free of non-perturbative QCD effects, so that a robust $\O(10^{-4})$ precision is achievable. However, it should be noted that this method assumes the validity of the SM, but the Z decay ratios may in general be modified by BSM physics. Similar considerations apply to the determination of $\alphas$ from the leptonic branching fraction of W bosons. For more information about future determinations of $\alphas$, see Ref.~\cite{dEnterria:2022hzv}.

\label{deltaalpha2}
As mentioned in the previous subsection, the shift $\Delta\alpha$ between the $\alpha(\mz)$ and $\alpha(0)$ is also an important ingredient for precision electroweak studies. Future $e^+e^-$ Higgs factories could in principle provide data for the dispersive approach using the radiative return method, $e^+e^- \to {\rm had.}+n\gamma$. While no detailed studies have been performed, it is not expected that this will lead to an improvement compared to data from lower-energy $e^+e^-$ colliders. On the other hand, with sufficient amounts of luminosity spent at two center-of-mass energy a few GeV below and above the Z peak, it is possible to determine $\alpha(\mz)$ directly, since the $\gamma$--$Z$ interference contribution is sensitive to this quantity \cite{Janot:2015gjr}. However, this method crucially depends on multi-loop theory calculations for the process $e^+e^- \to \mu^+\mu^-$.

\begin{figure}[t]
    \centering
    \includegraphics[width=0.65\textwidth]{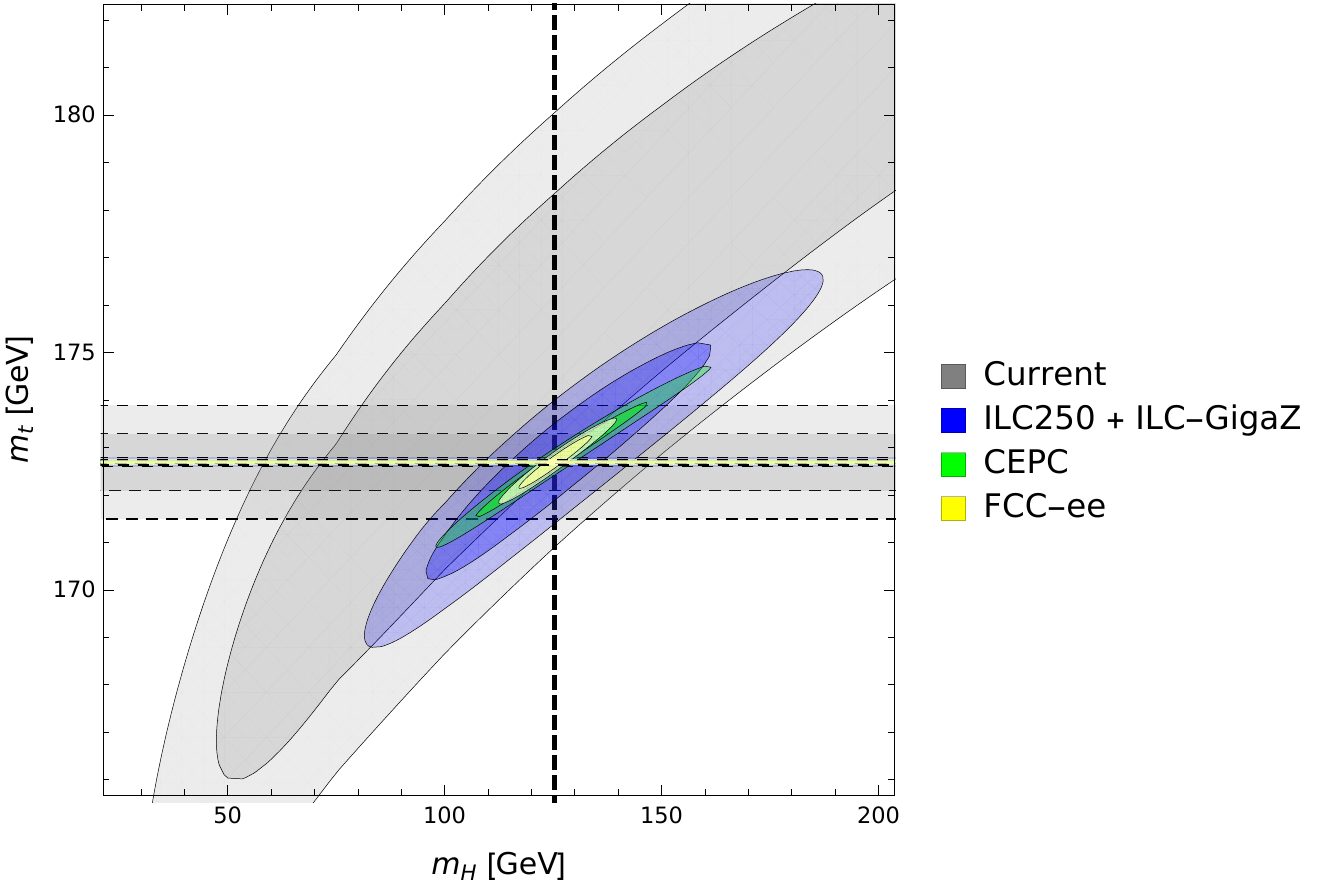}
    \medskip
    \caption{Indirect sensitivity to $\mh$ and $\mt$ for a fit of SM theory predictions to current and projected future data for electroweak precision tests (W mass and Z-pole quantities). For comparison, the direct measurement precision is also shown (on the scale of the plot the width of the $\mh$ band is not visible). The light (dark) shaded areas depict 95\% (68\%) confidence level regions. For the future collider scenarios it is assumed that the central values coincide with the SM expectations. The plot was made using SM theory predictions from Refs.~\cite{Awramik:2003rn,Dubovyk:2019szj}.}
    \label{fig:efmain:FitSM-EW}
\end{figure}

Since the list of EWPOs in Tab.~\ref{tab:ewpocomp} is an over-constrained set of inputs for a SM fit, it can be used to indirectly determine the Higgs-boson and top-quark masses, which only appear within loop corrections. This is illustrated in Fig.~\ref{fig:efmain:FitSM-EW}, which demonstrates that all future $e^+e^-$ colliders will tremendously improve the precision of this indirect test compared to the currently available data. The increased precision for the indirect determination of $\mh$ and $\mt$ at CEPC/FCC-ee compared to ILC is driven by the higher expected precision for the EWPOs themselves and for the strong coupling constant $\alphas$. For ILC we assume $\Delta\alphas = 0.0005$, while for CEPC/FCC-ee we use $\Delta\alphas = 0.0002$ \cite{dEnterria:2022hzv}. The difference in contours between CEPC and FCC-ee is mostly due to different assumptions about the precision of $\alpha(\mz)$, where for FCC-ee we consider the direct determination according to Ref.~\cite{Janot:2015gjr} with $\Delta\alpha(\mz) \sim 3 \times 10^{-5}$. On the other hand, we take the present-day uncertainty $\Delta\alpha(\mz) \sim 1 \times 10^{-4}$ for CEPC, which is excessively consersative but serves to illustrate the impact of $\Delta\alpha(\mz)$ in the electroweak precision fit.

%-------------------------------------------------------------------------
%-------------------------------------------------------------------------

\paragraph{Further improvements:} Some of the statistical and systematic uncertainties discussed above may be further improved with new advances of detector design and reconstruction techniques. Ref.~\cite{Yumino:2022vqt} studied the prospects for precision measurements of the tau polarization, which can be used for measuring Z-boson coupling ratios, as mentioned in section~\ref{sec:ewpocurr}. Due to the unobserved neutrino in the tau decays, a direct measurement of the tau polarization is not possible. Approximate polarization observables can be defined using only the visible decay products of individual tau decays. A better approximation of the true polarization may be obtained by using the full visible event information in di-tau events, $e^+e^- \to \tau^+\tau^-$. In this case it is possible to kinematically reconstruct the invisible neutrino momenta up to a two-fold ambiguity. However, all of the aforementioned polarization measurement methods require knowledge of the collision energy, and thus they suffer from ISR and beamstrahlung. 

A new reconstruction method, which is much less sensitive to ISR, also makes use of the impact parameter of the visible tau decay products with the beam axis \cite{Yumino:2022vqt}. This information allows one to reconstruct the tau momenta exactly in the presence of a single ISR photon. For tau decays, the observable impact parameters are typically below 1 mm and thus they require precise vertex detectors. The impact parameter method appears promising to achieve 70--80\% efficiency for the tau momentum and polarization reconstruction \cite{Yumino:2022vqt}.

\subsection{Electroweak precision measurements at other facilities}

Besides high-energy $pp$ and $e^+e^-$ colliders, other experiments can also perform interesting precision measurements of the electroweak sector.

At electron-positron colliders with $\sqrt{s} \ll \mz$, the process $e^+e^- \to f\bar{f}$ is dominated by photon exchange, and thus it is less sensitive to electroweak physics. However,
the Belle II experiment at the SuperKEKB collider with $\sqrt{s} = 10.58\,\GeV$ will benefit from the very large integrated luminosity to obtain some competitive constraints. In particular, an upgrade SuperKEKB with polarized electron beams would open up the possibility of measuring the left-right asymmetry of the process $e^+e^- \to f\bar{f}$ \cite{BelleSnowmass}. For $\sqrt{s} \ll \mz$, this process is mainly sensitive to $v_f$, the vector coupling of the Z-boson to $f\bar{f}$. With 40~ab$^{-1}$ integrated luminosity, the precision of $v_f$ for $f=\mu,b,c$ could be improved by a factor of 4--7 compared to the current world average. For most final states, the precision is statistics limited, except for the $b\bar{b}$ final state. The dominant systematic error sources are the polarization measurement (0.3\%) and subtraction of backgrounds (which include the $\Upsilon$ resonances) \cite{BelleSnowmass}.

One also can interpret the results for $v_f$ as a determination of the running weak mixing angle in the $\MSB$ scheme, $\sin^2\bar\theta(\mu)$. The achievable precision is comparable to the combined LEP+SLC precision, but at a lower scale $\mu \approx 10\,\GeV$, thus providing a non-trivial test of the running of $\sin^2\bar\theta(\mu)$.

Similarly, various low-energy precision experiments can determine the running weak mixing angle at very small scales, $\mu \lesssim 1\,\GeV$, through measurements of parity violation in fixed-target electron scattering and in atomic physics, see Ref.~\cite{Erler:2021eyn} for a brief overview.

On the other hand, new information on the running of $\sin^2\bar\theta(\mu)$ at larger scales, $10\,\GeV < \mu < 60\,\GeV$ will be accessible at the Electron-Ion Collider (EIC), using scattering of electron and positron beams on proton and deuteron beams \cite{Zhao:2016rfu,Boughezal:2022pmb}.

While $e^+e^-$ colliders can deliver the best precision for many EWPOs, it is difficult to disentangle individual couplings of gauge bosons to light quarks, due to the low sensitivity of tagging light-quark flavor and charge.
Lepton-proton colliders are ideally suited to overcome this difficulty. With the possibility of switching between $e^-p$ and $e^+p$ runs and with polarized $e^+$ beams, it is possible to individually determine the vector and axial-vector couplings of the Z-boson to light quarks ($v_d$, $a_d$, $v_u$, $a_u$) and simulaneously constrain the relevant PDFs. This is achieved by measuring neutral-current deep-inelastic scattering (DIS), $e^\pm p \to e^\pm +X$, where photons and Z-bosons appear in the t-channel. The relatively contribution of Z-boson increases with higher energies, so that a future high-energy $ep$ collider will have substantially higher sensitivity to these couplings than previous experiments.

\begin{figure}[tb]
\centering
\includegraphics[width=2.5in]{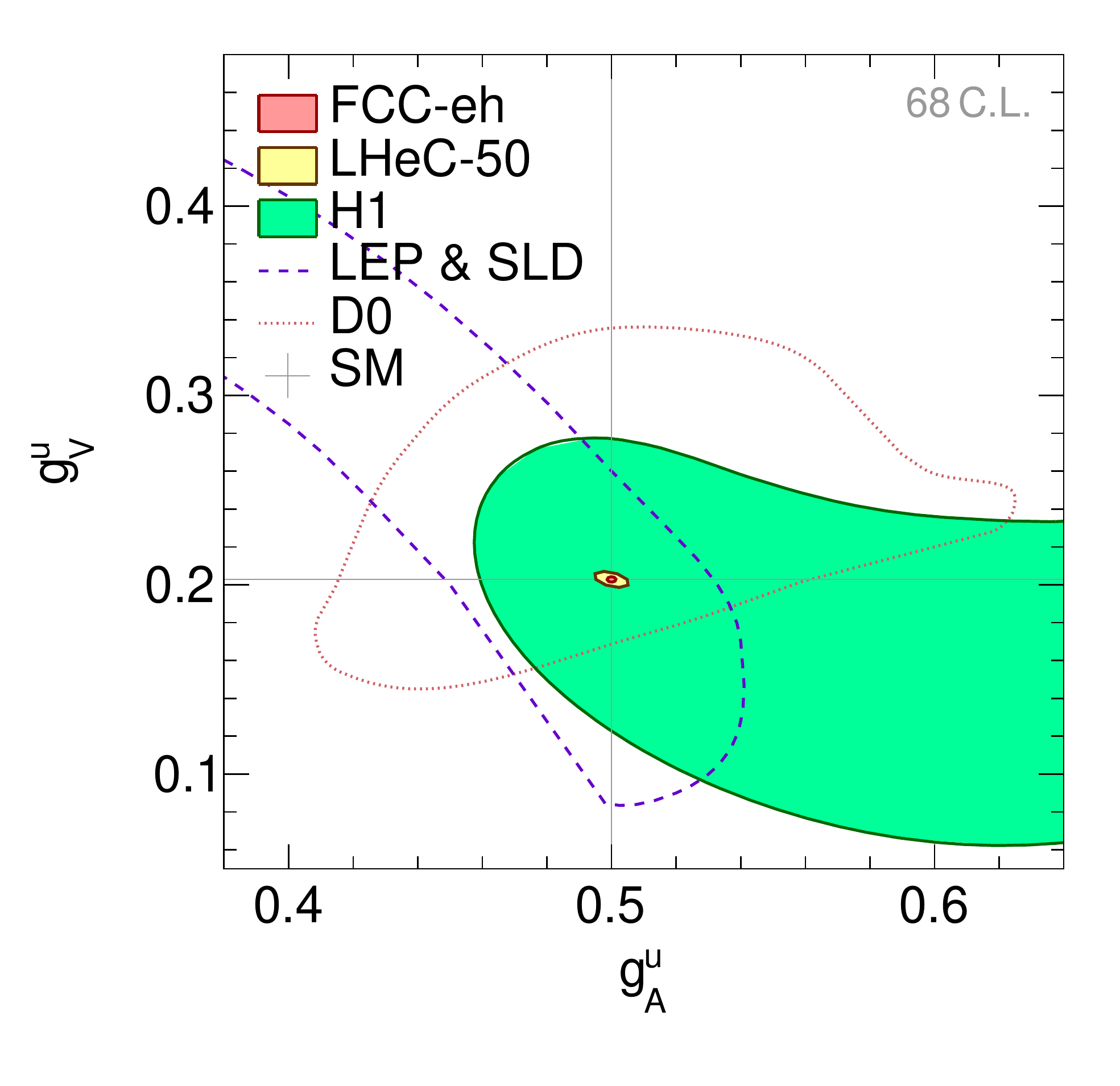}
\includegraphics[width=2.5in]{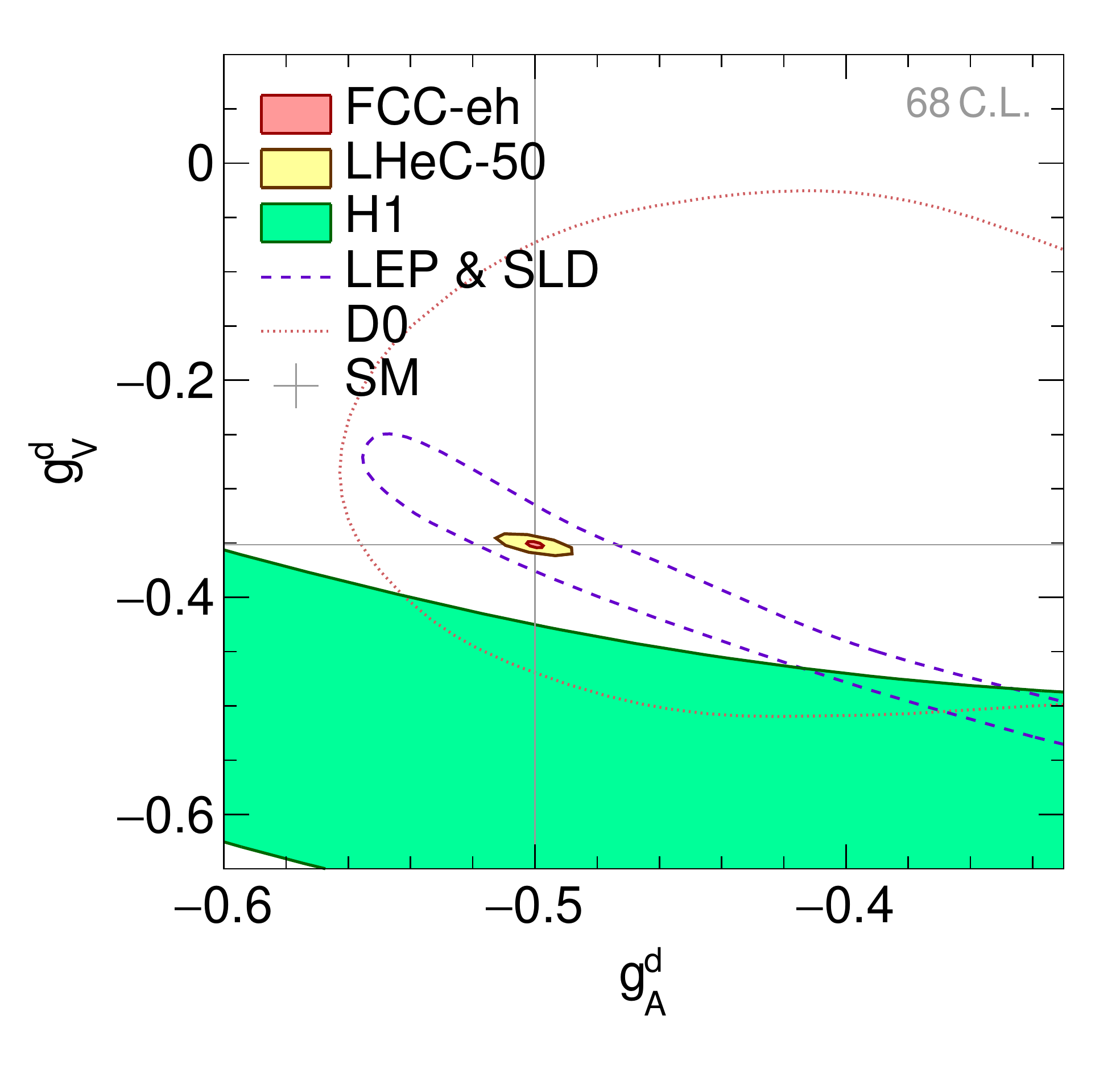}
\vspace{-1em}
\caption{Projected precision for the Z-boson vector and axial-vector couplings to light quarks at LHeC and FCC-eh, compared to the current precison from LEP, SLC, TeVatron and HERA (figure taken from Ref.~\cite{Britzger:2022abi}). Here $g_V^f$ and $g_A^f$ are rescaled versions of the vector and axial-vector couplings introduced in eqs.~\eqref{eq:Gpar},\eqref{eq:Hpar}: $v_f = g/(2 \cos\theta_w)\,g_V^f$, $a_f = g/(2 \cos\theta_w)\,g_A^f$.}
\label{fig:EWDIS}
\end{figure}

Two proposals for such a collider utilize the proton beam from the LHC (called LHeC) \cite{LHeC:2020van} or from FCC (called FCC-eh) \cite{FCC:2018byv}, with center-of-mass energies of 1.3~TeV and 3.5~TeV, respectively. By performing a simultaneous coupling and PDF fit, it was found the LHeC can determineall for couplings ($v_d$, $a_d$, $v_u$, $a_u$) with $\O(\%)$ precision \cite{Britzger:2020kgg}, while the precision can be improved by another factor 2--3 at FCC-eh \cite{Britzger:2022abi}, see Fig.~\ref{fig:EWDIS}.
%-------------------------------------------------------------------------
%-------------------------------------------------------------------------

A muon collider with center-of-mass energy $\sqrt{s} \approx 91$~GeV \cite{Blondel:1997eq} could perform electroweak measurements with a precision that greatly exceeds that currently available data from LEP/SLC. More studies for electroweak physics at muon colliders would be important to more thoroughly assess its potential.

\subsection{Theory needs for the interpretation of Electroweak precision data}

To fully exploit the potential of electroweak precision measurements to test the SM and possible new physics effects, theory inputs are needed in multiple places:
\begin{itemize}
    \item Most of the quantities in Tab.~\ref{tab:ewpocomp} are not real observables, but \emph{pseudo-observables}. The pseudo-observables are defined without backgrounds, initial-state radiation (ISR), the impact of final-state QED/QCD radiation on distributions, and detector smearing and acceptance effects. Various corrections factors and subtraction terms are needed to translate real observables to pseudo-observables. While it is possible to extract some of these terms with data-driven methods, theory input is needed in many instances, either for calibration or because the data-driven methods do not capture all relevant effects. The current state of the art are NLO results for the irreducible background contributions, MC tools with full NLO and partial higher-order QED radiation, and higher-order initial-state photon radiation in a leading-log approximation (see \eg\ Ref.~\cite{Freitas:2020kcn} for a review). For the expected precision of future $e^+e^-$ Higgs factories, one more order of perturbation theory (NNLO) will likely be needed for the background contributions, and one or more orders of improvement are required for the simulation of QED effects in MC tools, which may require novel frameworks for the architecture of MC programs \cite{Jadach:2019bye,Jadach:2019wol,Frixione:2022ofv}.
    \item For the interpretation of measured values of the pseudo-observables, they need to be compared to precise predictions within the SM. For Z-pole EWPOs, full NNLO and partial higher-order corrections are currently known, while NLO plus partial higher orders are available for most other processes (such as $e^+e^- \to WW$). The estimated theory uncertainties are subdominant compared to current experimental accuracies, but are significantly larger than the anticipated precision of future $e^+e^-$ collider, cf.\ Tabs.~\ref{tab:ewpocurr}, \ref{tab:ewpocomp}, \ref{tab:ewpoth}. The dominant missing contributions are 3-loop corrections with at least one closed fermion loop\footnote{Corrections with fermion loops are enhanced due to the large top Yukawa coupling and the large fermion multiplicity in the SM.} and leading 4-loop corrections enhanced by powers of the top Yukawa coupling \cite{Freitas:2019bre}. It is not possible to provide a reliable projection for how much the availability of these corrections would reduce the overall theory uncertainty, but a very rough estimate has been attempted in Ref.~\cite{Freitas:2019bre}, using a combination of methods (extrapolation of the perturbation series, counting of known prefactors, scheme comparisons). As shown in Tab.~\ref{tab:ewpoth}, these corrections will likely be needed to match the precision of future $e^+e^-$ colliders (Tab.~\ref{tab:ewpocomp}), and in some cases even higher orders may be necessary. Fortunately, there is continuous progress in the development of new calculational techniques for loop diagrams \cite{Liu:2021wks,Liu:2022chg,Song:2021vru,Dubovyk:2022frj}.
%-------------------------------------------------------------------------
\begin{table}[tb]
  \begin{center}
  \begin{tabular}{|c||c|c||c|c|c|}
    \hline
    EWPO & Current & Projected & Current & \multicolumn{2}{c|}{Projected param.\ error} \\
    \cline{5-6}
    uncertainties        &    theory error     &  theory error & param.\ error & Scenario 1 & Scenario 2    \\
    \hline
    $\Delta\mw$ (MeV)          &    4   &    1   & 5  & 2.8 & 0.6  \\
    $\Delta\gz$ (MeV)     &    0.4    &  0.1  & 0.5 & 0.3 & 0.1 \\ 
    $\Delta\sin^2\theta^\ell_\eff\; (\times 10^{5})$   &    4.5  & 1.5 & 4.2 & 3.7 & 1.1   \\ 
    $\Delta A_\ell\; (\times 10^{5})$   &  32 &  11  &  30 & 25 & 7.5   \\ 
    $\delta R_\ell\;(\times10^3)$  &    6    &    1.5   &  6 &  3.2 & 1.3  \\
    \hline    
  \end{tabular}
  \end{center}
\vspace{-2ex}
\caption{Impact of theory and parametric uncertainties on the prediction of a few selected EWPOs (see Ref.~\cite{Freitas:2019bre}). For the theory errors, the uncertainty estimates from currently available calculations are compared to the projected improvement 
when assuming the availability of N$^3$LO corrections and leading N$^4$LO corrections. For the parametric errors, current uncertainties are compared to two future scenarios, see eq.~\eqref{eq:parscen}.}
\label{tab:ewpoth}
\end{table}
%-------------------------------------------------------------------------
\item Furthermore, as already mentioned above, SM theory predictions of EWPOs require various SM parameters as inputs, most notably the top mass $\mt$ and Higgs mass $\mh$, the strong coupling constant $\alphas$, and the shift of the fine structure constant, $\Delta\alpha$. While the latter has been discussed above on pages \pageref{deltaalpha1} and \pageref{deltaalpha2}, information about the other parameters can be found in Ref.~\cite{dEnterria:2022hzv} and the EF Higgs and TOPHF reports \cite{Dawson:2022zbb,Schwienhorst:2022hht}.

The impact of SM parameter uncertainties are illustrated in Tab.~\ref{tab:ewpoth} for current results for these parameters and two future scenarios:
\begin{equation}
\begin{tabular}{|l|ccccc|}
\hline
 & $\Delta\mt$ [GeV] & $\Delta\mh$ [GeV] & $\Delta\mz$ [MeV] & $\Delta(\Delta\alpha)$ & $\Delta\alphas$ \\
\hline
Current & 0.6 & 0.17 & 2.1 & $10^{-4}$ & $9\times 10^{-4}$ \\
Scenario 1 & 0.3 & 0.02 & 0.8 & $10^{-4}$ & $5\times 10^{-4}$ \\
Scenario 2 & 0.05 & 0.01 & 0.1 & $3 \times 10^{-5}$ & $2\times 10^{-4}$ \\
\hline
\end{tabular}
\label{eq:parscen}
\end{equation}
Scenario 1 approximately corresponds to a Higgs factory with a Giga-Z Z-pole run and no data taking at the $t\bar{t}$ threshold. Scenario is more similar to a Higgs factory with a Tera-Z Z-pole run (FCC-ee, CEPC) and including $t\bar{t}$ threshold run. In particular, the improvement in the $\mt$ precision is crucial for reducing the parametric uncertainties in Scenario 2, to a level that is roughly comparable to the target precision for these EWPOs shown in Tab.~\ref{tab:ewpocomp}.

Note that the dependence of the predictions for $\gz$ and $R_\ell$ on $\alphas$ are to a certain extent circular, since these quantities would be used for the extraction of the strong coupling constant at future $e^+e^-$ colliders \cite{dEnterria:2022hzv}.
\end{itemize}

%%%%%%%%%%%%%%%%%%%%%%%%%%%%%%%%%%%%%%%%%%%%%%%%%%%%%%%%%%%%%%%%%%%%%%%%%%%%
%%%%%%%%%%%%%%% Multi-boson

\section{Multi-boson processes at high-energy colliders}
\label{multbos}

The SM predicts the existence of multi-boson interactions, which give rise to final states with two or three bosons. Anomalies in the rate and kinematic of these final states can be indicative of new physics not currently described in the SM. Such anomalies can be parametrized through modifications of the strength or form of the SM multi-boson vertices. A newer approach consists in using EFT operators of dimension six or above, where measurements of multi-boson processes can be recast as direct determinations of the Wilson coefficients of these operators.

It shall be noted that the sensitivity to BSM effects, or, in other terms, the upper limits to the Wilson coefficients of new operators, scale with a power of the c.o.m.\ energy, thus making multi-TeV colliders the ideal tools for studying these final states. At this time, the most promising avenues for reaching multi-TeV energies are proton-proton colliders or $\mu^+\mu^-$ colliders.

Di-boson final states can be produced directly through annihilation of the colliding particles or partons, or indirectly through vector-boson fusion (VBF) / vector-boson scattering (VBS) processes. These processes can give important clues about the origin of the electroweak symmetry breaking, and whether the Higgs mechanism is the only source of it. Hadron colliders also offer the possibility to study same-sign WW production through VBS. Other interesting final states contain three bosons, such as WWW, or the as-of-yet unobserved ZZZ. These final states can be produced via quartic-gauge couplings, and allow one to unveil one of the ostensibly least known sector of the SM.

As noted above, discrepancies between the expected total and differential cross sections for each
of the multi-boson final states and their SM predictions can be studied with two
different approaches. A modification of the strength of the SM couplings constitute the
premise of the searches for triple- and quartic-gauge-coupling anomalies (TGC and QGC, respectively).
The fundamental assumption is that there are no additional types of interaction among SM particles than
the ones already included in the SM Lagrangian. The adoption of EFT operators allows one
to eliminate this constraint, and bestows the freedom to obtain a model-independent
extension of the SM Lagrangian, under the assumption that there are no additional fields. The SMEFT approach is described in more detail
in Section~\ref{subsec:eftframeworkscope}.

Plentiful experimental results with multi-boson final states are available.
Both the ATLAS and CMS collaborations have measured di-boson \cite{Aaboud:2019lgy,Aaboud:2019nkz,Aaboud:2019gxl,Aaboud:2018jst,Aaboud:2017rwm,Sirunyan:2019gkh,Sirunyan:2019ksz,Sirunyan:2019bez,Sirunyan:2017zjc}, tri-boson processes \cite{Aad:2019dxu,Aaboud:2017tcq,Sirunyan:2020cjp,Sirunyan:2017lvq}, as well as VBF/VBS processes \cite{Aad:2020zbq,Aad:2019wpb,Aad:2019xxo,Aaboud:2019nmv,Aaboud:2018ddq,CMS:2020zly,Sirunyan:2020gyx,Sirunyan:2020tlu,Sirunyan:2019der,Sirunyan:2017fvv,Sirunyan:2018vkx}, which are characterized by a $VVjj$ final state. Di-boson final states include $W^+W^-$, same-sign $W^\pm W^\pm$, $WZ$, $ZZ$, $Z\gamma$. Tri-boson final states include $W\gamma\gamma$, $Z\gamma\gamma$, $WV\gamma$ (where $V=W,Z$), and $WVV'$ (where $V,V'=W,Z$). A summary of the expected sensitivities of multi-boson cross-section measurements for HL-LHC is reported in
Ref.~\cite{HLLHC}.

Bounds on new physics have been determined in the language of anomalous gauge-boson couplings (aGCs) \cite{Aaboud:2019lgy,Aaboud:2018jst,Aaboud:2017rwm,Sirunyan:2019gkh,Sirunyan:2017zjc} and effective operators \cite{Aaboud:2019nkz,Sirunyan:2020gyx,Sirunyan:2020tlu,Sirunyan:2019gkh,Sirunyan:2019der,Sirunyan:2019ksz,Sirunyan:2019bez,Sirunyan:2017fvv,Sirunyan:2017lvq}. The latter is theoretically preferred since it provides a consistent power counting and allows one to implement theoretical consistency constraints. In these studies, only one or two aGCs/operators are allowed to be non-zero at the same time, i.e., no full aGC/SMEFT analysis has been performed.

The most up-to-date limits on gauge-coupling anomalies are available at Refs.~\cite{cmspublic_tgc,atlaspublic_tgc}.
Expected limits at the end of the HL-LHC and HE-LHC runs are reported in Refs.~\cite{Azzi:2019yne,Grojean:2018dqj}.

\subsection{Theory studies on anomalous couplings}

It is well know that diverging from the SM predictions of the TGC and QGC causes the growth of
scattering amplitudes, up to the point at which unitarity is lost. Various methods have been implemented
in TGC and QGC searches to enforce the conservation of unitarity, and the consistency of the analyses.
A study of the different methods employed by experimental collaborations is presented, e.g., in Ref.~\cite{Garcia-Garcia:2019oig}. A new direction of research toward the imposition of constraints
dictated by the necessity that SMEFT admits a UV completion is explored in Ref.~\cite{Li:2021lpe}.
The conventional approach to the derivation of positivity bounds uses 2-to-2 scattering amplitudes, showing that one can obtain a set of homogeneous inequalities for the dim-8 Wilson coefficients.
The limit of this technique is that it requires one to consider scattering amplitudes between
arbitrarily superimposed particle states, which has not been done systematically. The new approach
draws a connection between the positivity bounds of EFT Wilson coefficients and the solution of a geometric problem, i.e., finding the extremal rays of a spectrahedron, built from the crossing symmetries and the SM symmetries of an interaction amplitude. The bounds obtained with new method are compared to the ones from the elastic positivity bounds and shown to be more stringent. A concise survey of the recent advances in constraining the SMEFT parameter space from the UV considerations can be found in Ref.~\cite{deRham:2022hpx}, section 2.5.

A review of the usage and potential pitfalls of SMEFT is presented in~\cite{Shepherd:2022rsg}. As indicated before
in this report, EFT is the leading tool employed to determine, in a model-independent way (with certain symmetry assumptions), the impact of SM measurements on new physics. Two energy regimes, in which SMEFT studies are currently performed, are identified and
separately discussed: resonant and near-threshold processes at low energy; distribution tails at high energy.
In the former case, it is possible to show that the number of ways in which SMEFT can contribute is finite; this
allows to identify combinations of Wilson coefficients as contributors to the process in question, and effectively
obtain a resummation of all orders in the SMEFT expansion that affect the process of interest. In the latter case,
it is not possible to consider the full EFT effect on a process, and one needs to cut the expansion at dimension 6.
As hinted above, one then must consider the effect of truncating the expansion by neglecting higher-order operators, such as dim-8 ones, as well as the limits of validity of the EFT approximation.
The choice of a method to fit any model data is also discussed. A global fit, where all measurements are
considered on equal footing, is ideal, but requires significant work to properly combine and compare the fit inputs.
A sequential fit is presented as a quicker alternative, in which intermediate fits are performed by adding measurements
divided in subsets, in order of decreasing precision. Directions for future progress are summarized at the end of 
the paper, and involve studies of SMEFT at higher order for on-shell and near-threshold observables, the adoption
by experiments of an error estimation scheme for high-energy observables, and the development of sequential fits toward
the definition of a fully-global fitting framework.

\subsection{Multi-boson processes at future lepton colliders}

While limited in energy reach compared to hadron colliders, lepton colliders with $\sqrt{s} \gtrsim 1$~TeV have advantages for measurement of vector-boson scattering (VBS), due to the well-defined initial state, complete coverage of final states, and the possibility to separate spin, isospin and CP quantum numbers. Particle flow algorithms enable very good particle ID (to reduce photon-induced background) and W/Z discrimination from hadronic decays \cite{Marshall:2012ry}.

An $e^+e^-$ collider like ILC or CLIC can cover energies of a few TeV, while a muon collider or more speculative proposals such as plasma wakefield accelerators (e.g., Ref~\cite{Benedetti:2022tvd}) can reach tens of TeV \cite{Shiltsev:2019rfl}. In the latter case, VBS can be described with good accuracy by factorizing the full process $\ell\ell \to VV\ell'\ell'$ into a $V'V' \to VV$ hard process and $V'$ radiation in the initial state described by electroweak PDFs \cite{Costantini:2020stv,Han:2020uid}.

Significant backgrounds arise from a number of processes (including $\ell^+\ell^- \to VV$ without VBS), but they can be reduced with suitable cuts or machine learning techniques, and they also become less important relative to the signal process when going to higher values of $\sqrt{s}$ \cite{Beyer:2006hx,Fleper:2016frz,Green:2017fjw}. However, the achievable constraints on SMEFT coefficients do not always improve when increased center-of-mass energy \cite{Costantini:2020stv}.

Reference~\cite{Costantini:2020stv} presents a review of how electroweak vector boson fusion/scattering
processes become the dominant production modes of vector bosons as the center-of-mass energy of a lepton
collider enters the few-TeV range. The size and growth of VBF cross sections for numerous SM and new physics
processes are investigated. The key observation is that $s$-channel production rates decrease, with collider
energy, as $1/s$, while VBF rates grow as $\log s$, eventually becoming the most dominant process. 

A comprehensive review of VBS processes at current and future colliders is presented in Ref.~\cite{BuarqueFranzosi:2021wrv}. This paper also discusses the importance of adopt the proper formalism to describe initial-and final-state radiation (electroweak parton distribution functions, and resummation of fragmentation functions, respectively). This latter topic is presented also in Ref.~\cite{Ruiz:2021tdt} for lepton colliders (and similarly, in the scenario of a high-energy hadron collider, in Ref.~\cite{Han:2022laq}, which is discussed later in section~\ref{multi-boson-hadron}).
Multi-TeV lepton colliders are effectively weak-boson colliders, which suggests that EW bosons should be treated as constituents of
high-energy leptons. The paper reviews the validity of W and Z parton distribution functions, investigates
power-law and logarithmic corrections that arise in the derivation of weak boson PDFs in the collinear limit,
and reports an implementation of the Effective W/Z  and Weizs\"aker-Williams approximations into the Monte Carlo
generator \verb+MadGraph_aMC@LNO+. The key question is how factorization and resummation work in the weak sector, and
how it differs from QED and perturbative QCD. This question is critically important, as in multi-TeV muon
colliders, and at 100~TeV hadron colliders, typical parton collisions satisfy the criteria for collinear factorization
of weak bosons. It will furthermore be important to extend the factorization framework to higher orders (see \eg\ Ref.~\cite{Bertone:2019hks}).

The future lepton colliders obviously offer the opportunity to
investigate, in detail, interesting experimental signatures.
Three such studies are presented below, and touch two specific
aspects of lepton colliders: the precision study of Higgs physics, and the unique advantage (high-energy, high-statistics, clean environment) offered by muon colliders as weak-boson colliders.

\begin{figure}[tb]
\centering
\includegraphics[height=2.1in]{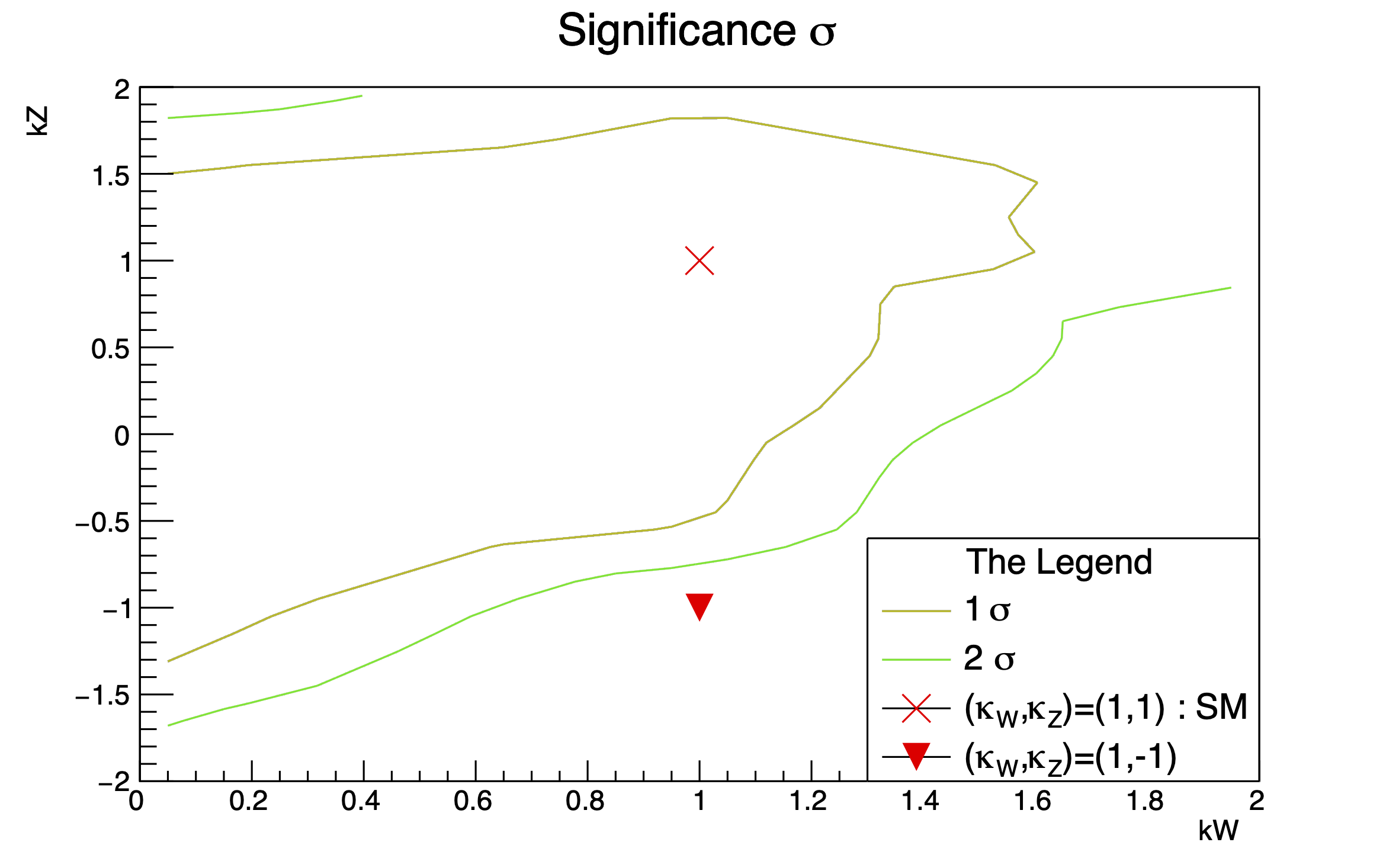}%
\includegraphics[height=2.1in]{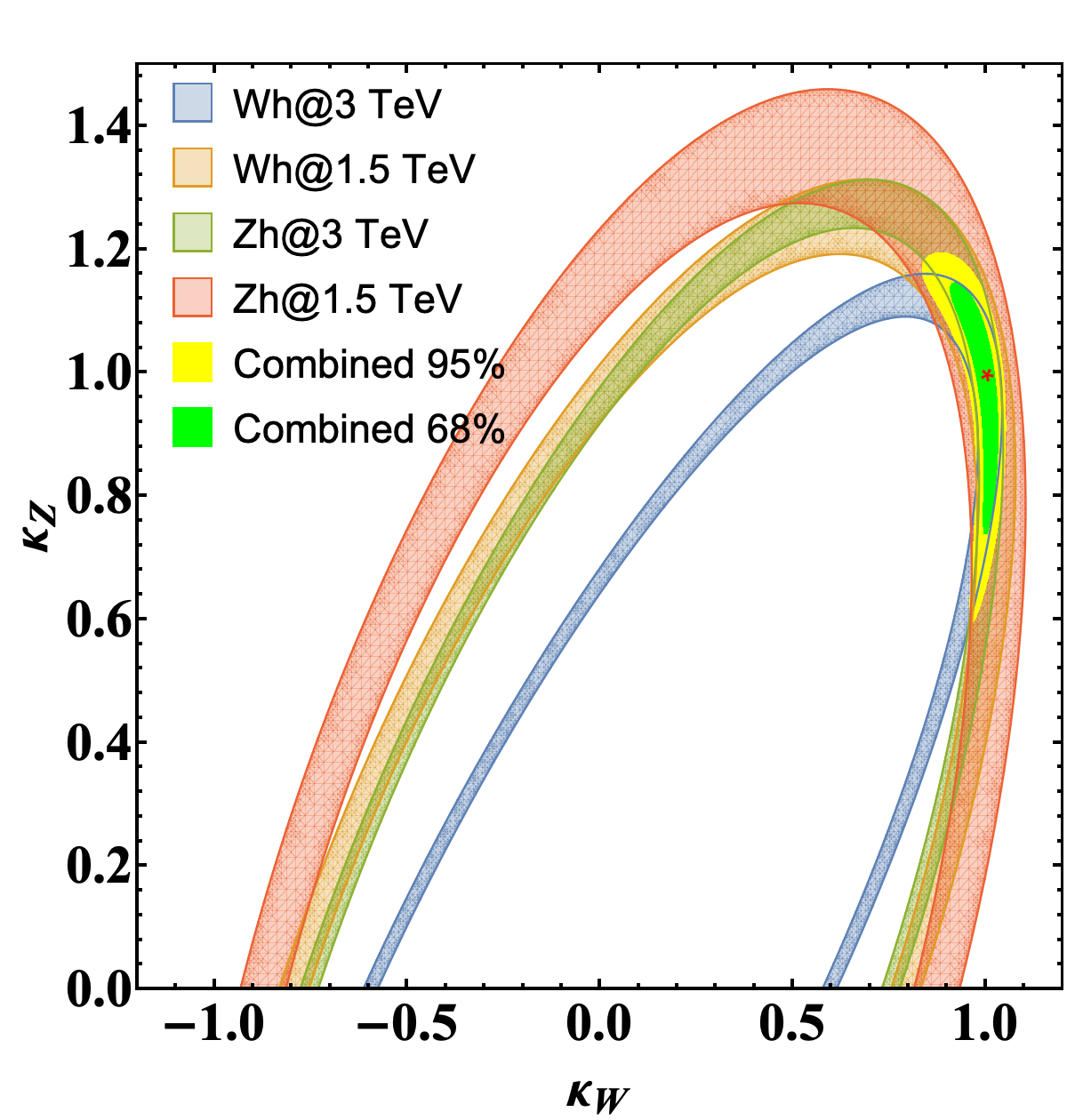}
\vspace{-1em}
\caption{Left: 1- and 2-$\sigma$ sensitivity of the measurement to $\kappa_W$ and $\kappa_Z$ at the HL-LHC. Right: constraints in the $\kappa_W-\kappa_Z$ plane for the total rate measurement at CLIC. (figures taken from Ref.~\cite{Paranjape:2022ekg}).}
\label{fig:lwz}
\end{figure}

Reference~\cite{Paranjape:2022ekg} reports a study of the Vh process that is relevant for both the HL-LHC
and future lepton colliders, in which the signal to background ratio is significantly higher than at the LHC.
A particularly interesting aspect of the analysis is the ability to check whether the Higgs couplings to W and Z, $\kappa_W$ and $\kappa_Z$ have the same sign; models in which they do not include scalars which have higher isospin representations.
The idea is to exploit the tree-level destructive interference between the W and Z mediated processes that contribute to the production of a Higgs boson in association with a vector boson via vector-boson fusion.
The Vh matrix element contains in fact a term that grows with energy and is proportional to $\lambda_{WZ}-1$,
where $\lambda_{WZ} = \kappa_W/\kappa_Z$ (i.e., $\lambda_{WZ}=1$ in the SM).
The future lepton collider being considered is CLIC, at 1.5~TeV and 3~TeV center-of-mass energy.
The achievable sensitivity at a lepton collider is shown in Fig.~\ref{fig:lwz}. It is reported that the point
$(\kappa_W,\kappa_Z)=(1,-1)$ is excluded at more than two standard deviation at the end of the HL-LHC run\footnote{Note that a more precise determination of the magnitude of $\kappa_{W,Z}$ can be achieved with a global fit of HL-LHC measurements, but the discussion here focuses only on the direct determination of the sign of these couplings.}, while
3.4 fb$^{-1}$ (14.1 fb$^{-1}$) are enough at CLIC 3 TeV (1.5 TeV) to exclude that point at 95\% CL against the SM case.

%-------------------------------------------------------------------------
%-------------------------------------------------------------------------

\begin{figure}[tb]
\centering
\includegraphics[width=2.5in]{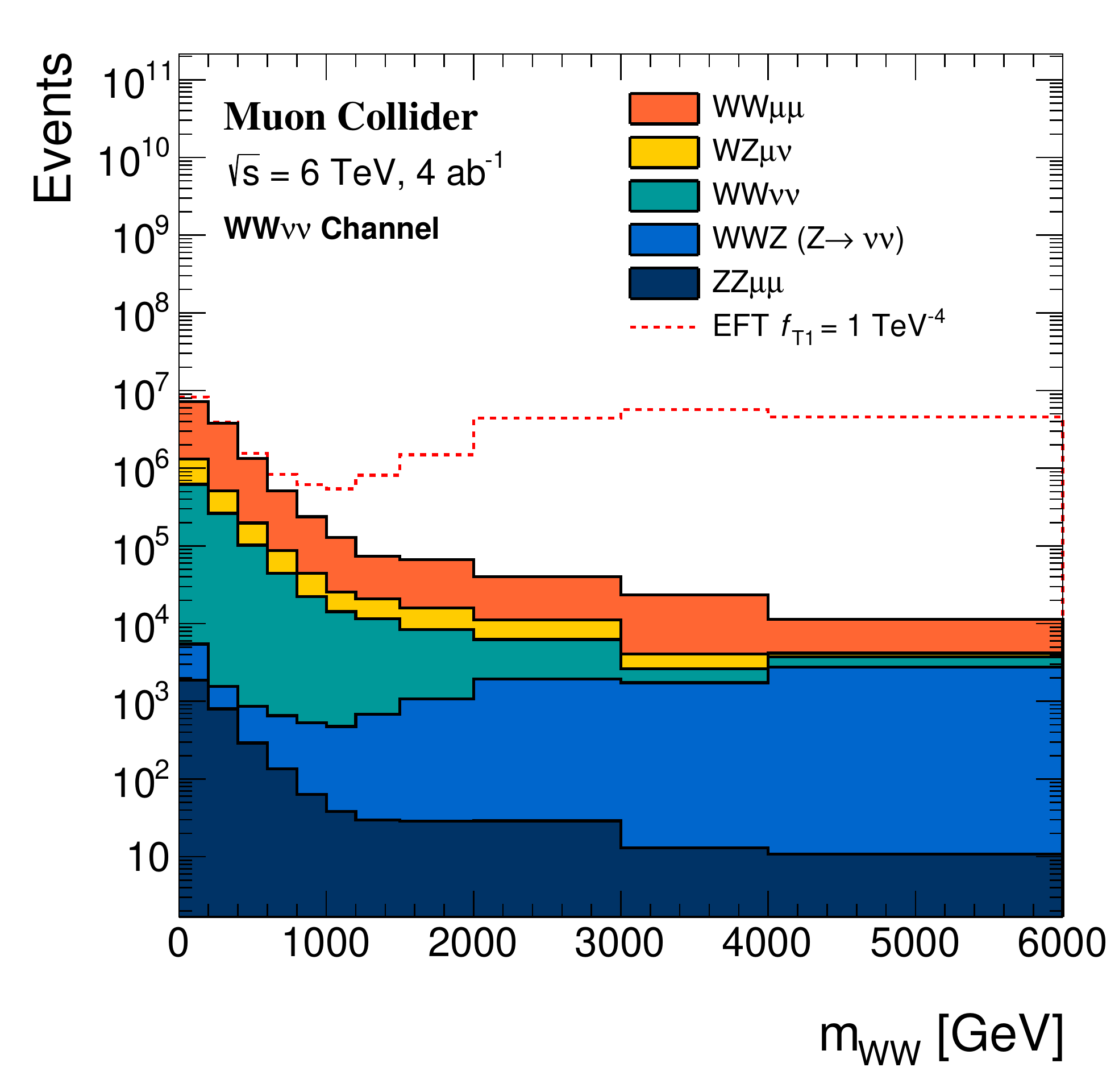}%
\includegraphics[width=2.5in]{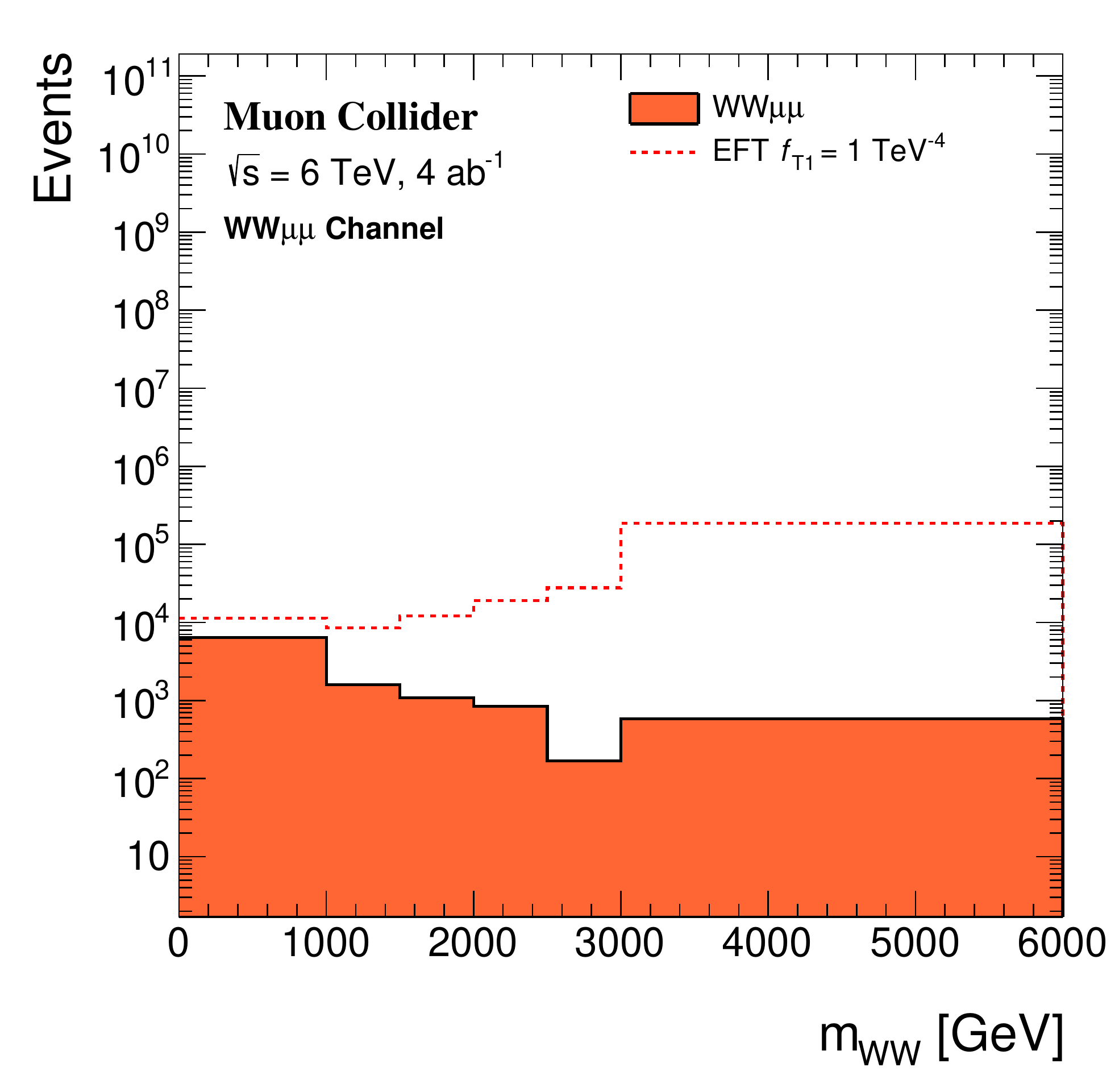}
\vspace{-1em}
\caption{Distribution of m(WW) in the WW$\nu\nu$ and WW$\mu\mu$ channels, after event selection. The dashed lines show the signal prediction for one illustrative aQGC parameter. (figure taken from Ref.~\cite{Abbott:2022jqq}).}
\label{fig:wwll_mww}
\end{figure}

Prospects for searches for anomalous quartic gauge couplings at a high-energy muon collider are presented
in Ref.~\cite{Abbott:2022jqq}. A multi-TeV muon collider is effectively a high-luminosity weak boson collider,
and allows for the measurement, in a relatively clean environment, of vector boson scattering processes.
The study of W pair production, in association with two muons or two neutrinos, is presented in the reference.
Deviations of the proposed measurements with respect to the SM predictions could indicate the presence of an
anomalous quartic gauge coupling. Figure~\ref{fig:wwll_mww} shows the distribution of the mass of the W pair
in the WW$\nu\nu$ and WW$\mu\mu$ channels, using a simulation with full matrix elements (rather than the effective W-boson approximation or EW PDFs). It also includes an example of a signal prediction for one illustrative aQCG parameter.
The limits on anomalous quartic gauge couplings that can be set with a luminosity of 4/ab of muon collisions at
a center-of-mass energy of 6~TeV are about two orders of magnitude tighter than the current limits. It shall be noted
that the effects of beam-induced background have not been included in the analysis, and that the limits on 
aQGCs are set under the assumption that triple gauge couplings are not modified.

%-------------------------------------------------------------------------
%-------------------------------------------------------------------------

Muon colliders also offers the opportunity to study in detail the topic of unitarity restoration, for example,
by measuring longitudinally polarized vector boson scattering. It is shown in Ref.~\cite{Yang:2021zak} that
such a study could surpass the end-of-life HL-LHC results in the ZZ channel. The study utilizes a Boosted Decision Tree and shows that, even with a conservative estimation, a 5 standard deviation discovery of longitudinally polarized
ZZ scattering can be achieved with 3/ab of data collected at a 14~TeV muon collider. This results outperforms
the expected results of the end-of-life HL-LHC, which expects to have a sensitivity of about 2 standard
deviations. The paper also reports the study of a 6~TeV muon-collider case, and shows that its sensitivity is
comparable to the HL-LHC one.

\subsection{Multi-boson processes at future hadron colliders}
\label{multi-boson-hadron}

Multi-TeV future hadron colliders provide a unique laboratory to explore the nature of EW symmetry breaking, and its restoration as energies increase above the EW scale.
The most compelling direction of investigation entails the study of vector bosons produced in association with a Higgs boson, or the longitudinal polarization of pair-produced vector bosons. Both
these lines of investigations are discussed below.

A discussion of the prospects for physics measurements at future hadron colliders requires a suitable description of the physics in the high-energy regime. 
Reference~\cite{Han:2022laq} discusses the definition of a consistent theoretical treatment
to describe the physics processes that take place in particle collisions at multi-TeV energies. 
In that regime, beyond the weak mass scale, all SM particles, including the gauge bosons, can be
considered to be massless. Collinear splitting becomes the dominant phenomenon. The proper description
of parton distribution functions, initial state radiations, final state radiations and fragmentation functions
is needed. The focus of the paper is the resummation of final-state radiation up to leading-log accuracy,
and show the effect of high-energy splitting at a 100~TeV collider. The formalism developed in the paper
is applicable also to the case of multi-TeV muon colliders.

\begin{figure}[tb]
\centering
\includegraphics[width=2.5in]{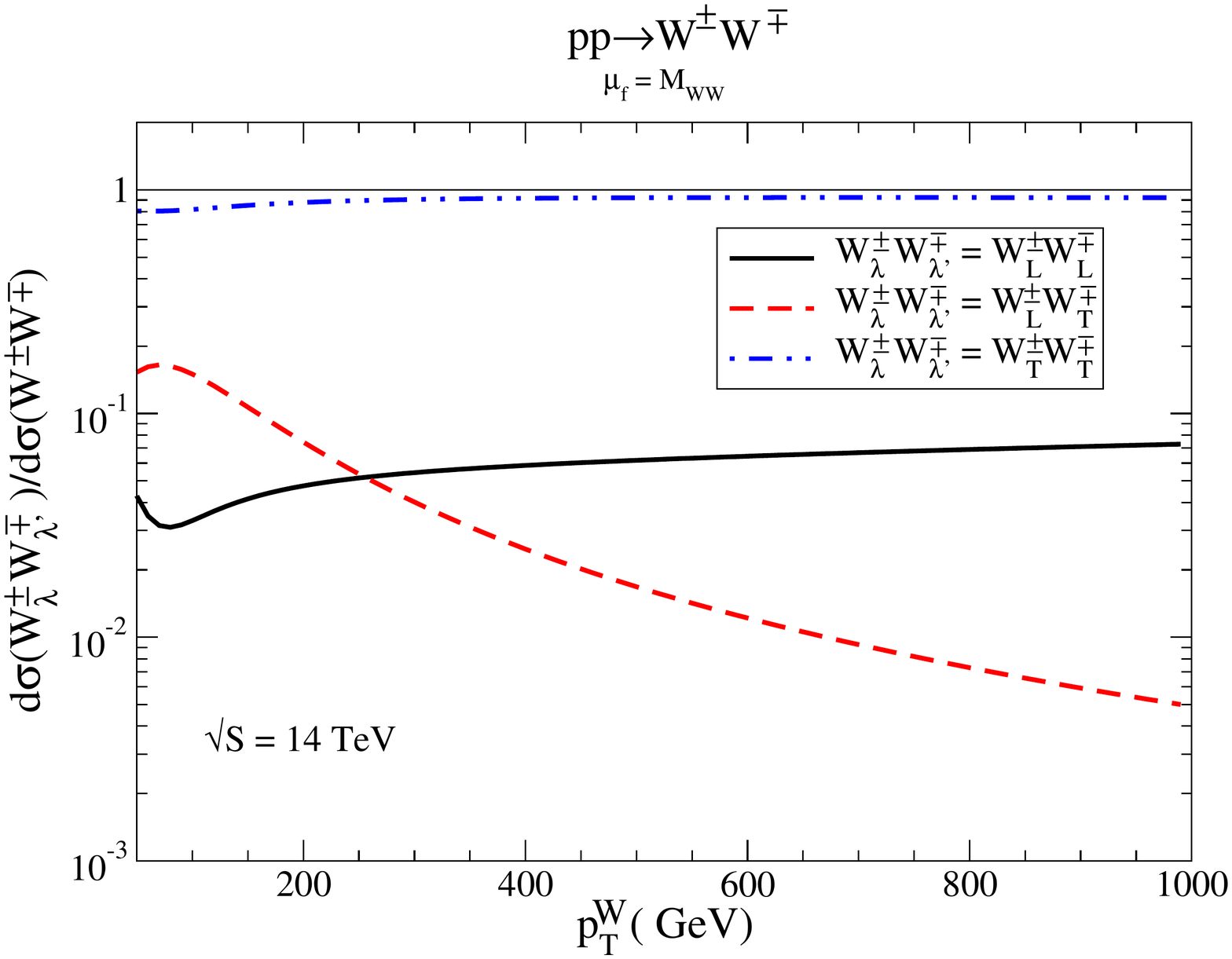}%
\includegraphics[width=2.5in]{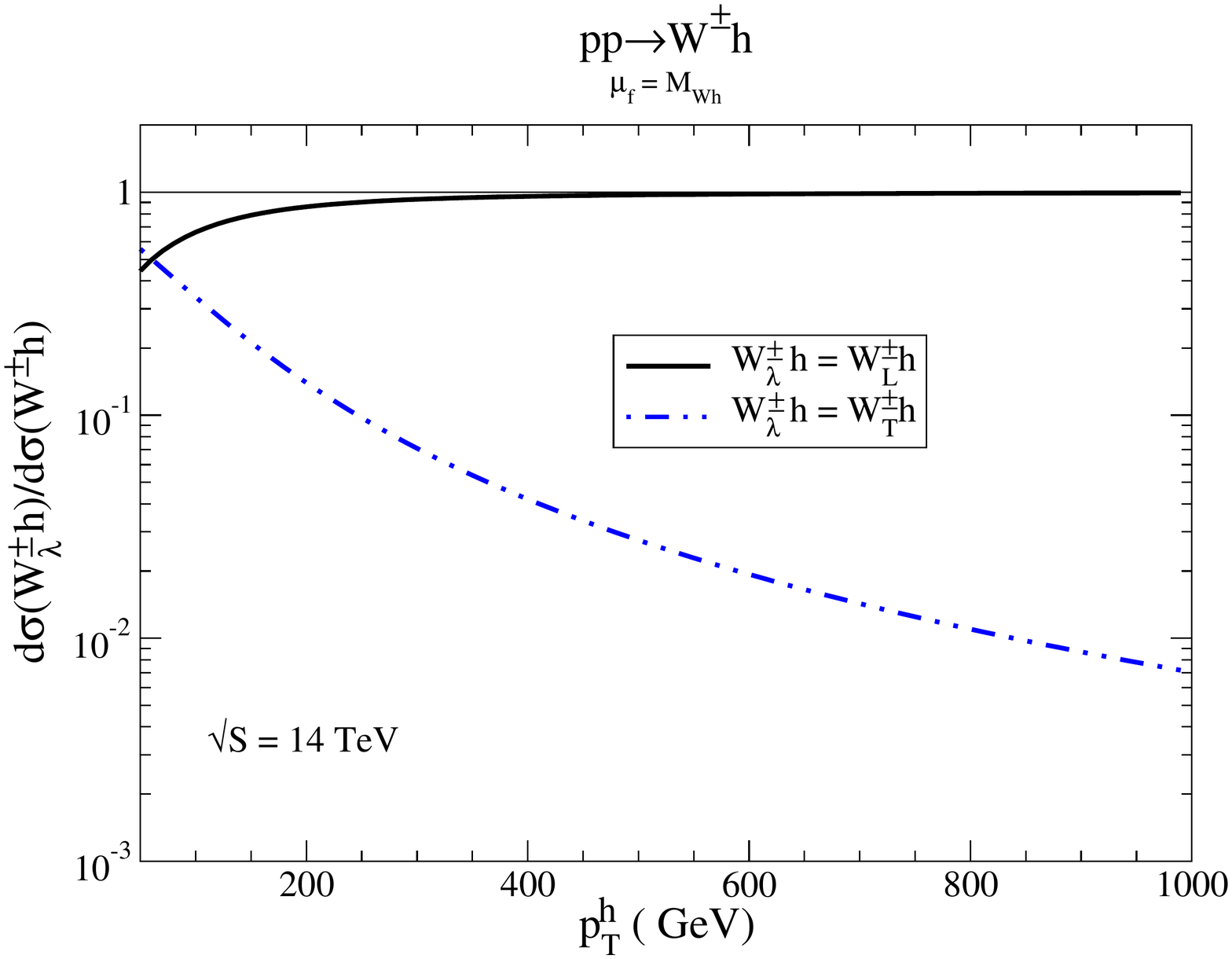}
\vspace{-1em}
\caption{Ratio of transverse momentum distributions of polarized gauge boson production to the
total distribution summed over polarizations. (figures taken from Ref.~\cite{Huang:2020iya}).}
\label{fig:whpol}
\end{figure}

As indicated earlier, the study of Higgs production in association
with a vector boson at a high-energy hadron collider offers a
test stand to probe the restoration of EW symmetry.
Reference~\cite{Huang:2020iya} presents a new test to study the restoration of EW symmetry at high energy. The two colliders under consideration are the 14~TeV HL-LHC and the 27~TeV HE-LHC. The
main assumption of the analysis is that at those energies the EW vector bosons become massless, and
one can replace their longitudinal modes with the associated Goldstone bosons. The conclusion
is that while the VV' production is dominantly transversely polarized, up to very high energies, the Vh channel is longitudinally dominated starting at relatively low energies (e.g., at a c.o.m. energy of 14~TeV, the W boson is longitudinally polarized in $\approx 90\%$ of the Wh events with a Higgs transverse momentum above 200~GeV). 
Figure~\ref{fig:whpol} demonstrates this effect, by showing the fraction of polarized gauge-boson production
as a function of the boson transverse momentum. It is ultimately demonstrated that the EW restoration can
be confirmed with a precision of 40\% at the HL-LHC, and 6\% at the HE-LHC.

%-------------------------------------------------------------------------
%-------------------------------------------------------------------------

\begin{figure}[tb]
\centering
\includegraphics[width=5in]{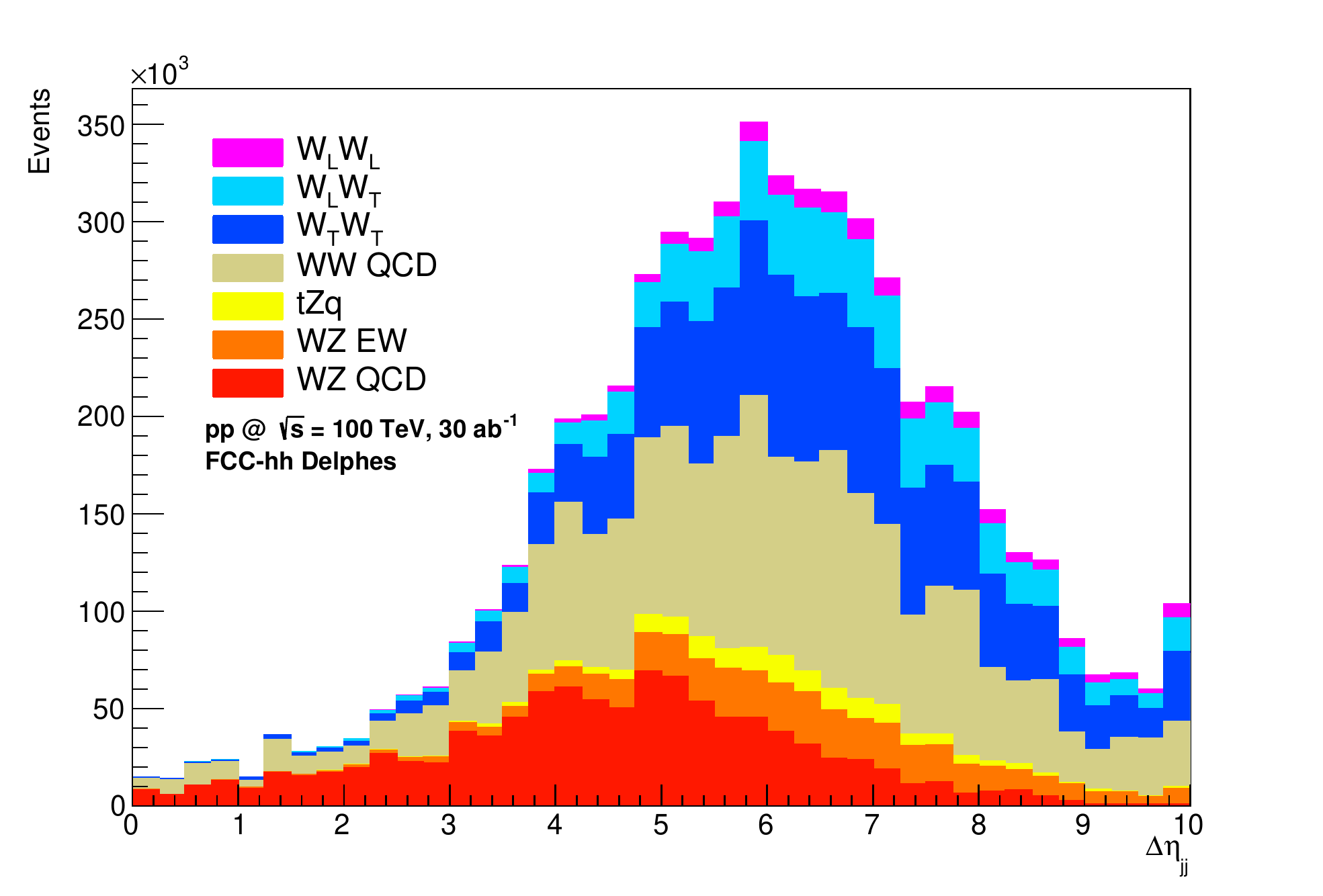}
\vspace{-1em}
\caption{The distribution of events as a function of $\Delta\eta(jj)$, the pseudorapidity difference between the two leading jets, after full event selection, for $\sqrt{s}=100$~TeV (figure taken from Ref.~\cite{Apyan:2022gis}).}
\label{fig:ssww_deta}
\end{figure}

The second compelling direction of investigation previously identified is the study of longitudinal polarization of pair-produced vector bosons.
The sensitivity to longitudinal vector boson scattering at a multi-TeV proton-proton collider, using
same-sign WW pairs produced in association with two jets, is presented in Ref.~\cite{Apyan:2022gis}.
Vector boson scattering processes are important probes of the non-Abelian structure of electroweak
interactions, as the unitarity of the tree-level amplitude of longitudinally
polarized gauge boson scattering could be restored at high energies by the Higgs boson. Extensions of the 
SM introduce new resonances or modifications of the Higgs boson couplings that modify the 
cross sections of processes involving the scattering of longitudinally polarized gauge bosons.
The same-sign WW state is of particular interest because the requirement of two same-sign leptons
in the final states greatly reduces the backgrounds from other SM processes. The study reported in the
paper assumes 30/ab collected at a center-of-mass energy of 27, 50, and 100~TeV. The result of the analysis
is that a cut-and-count method is sufficient to measure with a relative precision of 39\%, 22\%, and 17\% the
fraction of the purely longitudinal contribution to same-sign WWjj production, using the fully leptonic
decay mode (at a center-of-mass energy of 27, 50, and 100~TeV, respectively). The purely transverse and mixed longitudinal-transverse contributions are measured with a
relative precision of 2\% and 4\%, respectively, at 100~TeV. Figure~\ref{fig:ssww_deta} shows the distribution of the
signal ($W_LW_L$) and background components, as a function of the pseudorapidity difference between the two jets. As expected,
the signal-to-background ratio is higher at a large pseudorapidity difference.

%-------------------------------------------------------------------------
%-------------------------------------------------------------------------

%%%%%%%%%%%%%%%%%%%%%%%%%%%%%%%%%%%%%%%%%%%%%%%%%%%%%%%%%%%%%%%%%%%%%%%%%%%
%%%%%%%%%%%%%%%% Global fits

\section{Global fits of new physics}
\label{glob}

Assuming new physics scales are significantly higher than the EW scale, Effective Field Theories (EFT) provide a model-independent prescription that allows us to put generic constraints on new physics and to study and combine large sets of experimental data in a systematically improvable quantum field theory approach. 
All new physics effects are represented by a set of higher-dimensional operators which consist of only the SM fields and respect the SM gauge symmetries. 
Depending on whether the SU(2)$\times$U(1) gauge symmetries are realized linearly or nonlinearly, there are 
two classes of formalism popular for studying EW physics at colliders, 
the Standard Model Effective Field Theory (SMEFT)~\cite{Buchmuller:1985jz,Grzadkowski:2010es} or the Higgs Effective Field Theory (HEFT)~\cite{Buchalla:2013rka,Buchalla:2015wfa}; see Ref.~\cite{Brivio:2017vri} for a pedagogical review. 
The EFT approach has some features that are of particular interest for studying precision EW physics, for instance: it provides a well-defined theoretical framework that enables the inclusion of radiative corrections for both the SM and BSM parts; 
and the synergies between different precision EW measurements can be explored globally so that a comprehensive picture of the
constraints on new physics can be drawn.
However, the EFT approach also has some practical limitations since it has in principle an infinite number of degrees of freedom, and it is only an adequate description if the new physics scales are larger than the experimentally reachable energies. 
In a realistic global EFT fit, various flavor assumptions and truncations to the lowest order of 
relevant operators often have to be applied, to limit the number of parameters to a manageable level. 
The HEFT allows considerably more parameter freedom than the SMEFT. A subset of BSM physics that couples only to SM gauge bosons can also be represented by the so-called oblique parameters \cite{Peskin:1990zt,Altarelli:1990zd}. Two of the oblique parameters ($S$ and $T$) are directly related to dimension-6 SMEFT operators, while the $U$ parameter corresponds to a dimension-8 operator.

Most of the global EFT fits are currently performed based on the SMEFT, which will also be the focus here.
An up-to-date global SMEFT fit at future colliders has been performed for the European Study Group (ESG), which combines measurements of EWPOs, 
Higgs production and decay rates at LHC and future colliders, and $e^+e^-\to WW$~\cite{deBlas:2019rxi}. 
The results of global EFT fit may give important implications for proposals of future colliders which otherwise would not get recognized. Just to name two examples from ESG: $Z$-pole and $WW$ runs at circular $e^+e^-$ colliders can help improve significantly the Higgs coupling precisions with respect to what can be obtained using only $ZH$ runs; and beam polarization at linear $e^+e^-$ colliders can help lift degeneracies of different new physics effects, as a result of which similar Higgs coupling precision can be achieved at both linear and circular $e^+e^-$ colliders, in spite of the difference in integrated luminosity. 

For Snowmass 2021, the global EFT fit for ESG has been extended in a few directions~\cite{GlobalFitWhitePaper}: consistent implementation of full EFT treatment in $e^+e^-\to WW$ using optimal observables;
new inclusion of a large set of 4-fermion operators; more complete set of operators that are related to top-quark. The projections of the uncertainties of 
required input observables are provided by the Topical Group EF01 for Higgs related observables \cite{Dawson:2022zbb}, EF03 for top-quark related observables \cite{Schwienhorst:2022hht}, EF04 for $W/Z$ related observables (see section~\ref{ewprec}), and the
Rare Process and Precision Frontier (RF) for a set of low-energy measurements. 
%{\bf references to their reports?} 
The projections for various future $e^+e^-$ colliders are made to be as consistent as possible, for instance: by applying common systematic errors as explained in Section 2.2; by extrapolating from one collider to another whenever there is any important missing input. 
More details about the considerations of inputs can be found in Ref.~\cite{GlobalFitWhitePaper}. The global fits are performed with respect to various run scenarios for each collider. The results of global fits are provided as bounds on the Wilson coefficients as well as uncertainties of effective $H/Z/W$ couplings. 
The intrinsic and parametric theory errors of the SM predictions of observables are not included in the global fits by default, instead their impact is quantified separately in terms of how much change the theory error would bring on the 
fitted uncertainties on various effective couplings.
%[The parametric theory errors of which the prospects are known rather clearly are going to be included 
%in the default fits: for instance from uncertainties on Higgs mass, top mass, strong couplings, etc] {\bf TBC}. 
The computation of EFT contributions to various observables is done at tree level only except for
the loop contribution from the triple Higgs coupling in the single Higgs processes. 

\subsection{Framework and scope}
\label{subsec:eftframeworkscope}

The SMEFT takes a form of Effective Lagrangian from the SM part up to dimension-4 operators plus an infinite tower of higher-dimensional ($d>4$) operators ($O_i^d$)
which respect Lorentz and the SM gauge symmetries and are suppressed by the corresponding inverse powers of the cut-off scale $\Lambda$,
\begin{equation}
{\cal L}_{\rm SMEFT}={\cal L}_{\rm SM} + \sum_{d=5}^\infty \sum_i \frac{C_i^{(d)}}{\Lambda^{d-4}}{\cal O}_i^{(d)}. \label{eq:smeft}
\end{equation}
The information of new physics is encoded in the series of Wilson Coefficients $C_i^{(d)}$. 
The number of non-redundant operators at $d=5,6,7,8$ is known~\cite{Weinberg:1979sa,Wilczek:1979hc,Buchmuller:1985jz,Grzadkowski:2010es,Abbott:1980zj,Lehman:2014jma,Lehman:2015coa,Henning:2015alf}.
For the global fits presented here, we restrict ourselves to operators of dimension 6 ($d=6$) that preserve baryon and lepton numbers. 
A complete basis of such operators contains 2499 operators without flavor assumptions. 
It reduces to 84 if only one generation of fermions are considered, and further to 59 if only for CP-even operators. Three main global fits are performed for Snowmass 2021, called Fit-1/2/3, which each consider a different subset of operators to parametrize EW physics at future colliders. 
Independent operators for more than one generation of fermions are considered, with a general assumption
that the flavor structure is diagonal for simplicity, but without assuming lepton-flavor universality. The operators for the $3^{\rm rd}$-generation of quarks 
are always treated separately, while universality for the $1^{\rm st}$- and $2^{\rm nd}$-generation of quarks is assumed in Fit-1. 
Bounds on Wilson Coefficients are given in terms of the original Warsaw basis~\cite{Grzadkowski:2010es}. More details can be found in Ref.~\cite{GlobalFitWhitePaper}.

\begin{figure}%[t]
    \includegraphics[width=\textwidth]{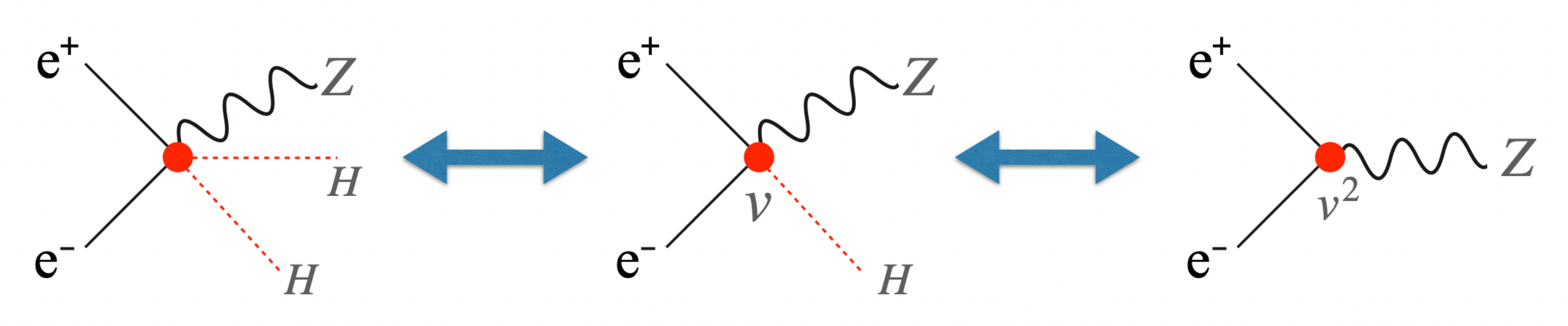}
    \caption{The contribution from the same operator ${\cal O}_{\phi e}$ (defined in the
    text) in three different processes: double Higgs production (left),
    single Higgs production (middle) and $Z$-pole production (right).}
    \label{fig:EFT_OP_cHE}
\end{figure}

Fit-1 is mainly focused on the Higgs and EW sectos. It explores the interplay among measurements for Higgs production and decay rate, EWPOs and di-boson processes; the roles played by energy, luminosity and 
beam polarizations; and the synergy between LHC and future lepton colliders. 
There are around 20 operators that contribute to those measurements, which is the complete set
given the assumptions as mentioned above. Let us consider
one of the operators as an example to illustrate why Higgs and EW measurements are inherently related: ${\cal O}_{\phi e} = (\phi^{\dagger}i\overset{\leftrightarrow}{D}_\mu \phi)
      (\bar{e}_R \gamma^\mu e_R)$. 
As sketched in Fig.~\ref{fig:EFT_OP_cHE}, this operator will directly generate a five-point interaction that contributes
to $e^+e^-\to ZHH$. By replacing the Higgs field by $v$, it will also generate
a four-point interaction that contributes to $e^+e^-\to ZH$ which is one of 
the leading Higgs production channels. Furthermore, replacing the other Higgs
field by $v$, it will result in a vertex correction to $Z$ pole observables. 
Therefore, the interplay of Higgs measurements and EWPOs at the $Z$-pole
together will be advantageous for probing the effects from new physics. 

Fit-1 results are given mainly in terms of effective couplings~\cite{Barklow:2017suo,deBlas:2019rxi} which are defined by pseudo-observables and thus are independent of the operator basis one would have chosen for the fit. 
The Higgs effective couplings ($g_{HX}^{\mathrm{eff}~2}$) are defined as 
\begin{equation}
g_{HX}^{\mathrm{eff}~2}\equiv \frac{\Gamma_{H\to X}}{\Gamma_{H\to X}^{\mathrm{SM}}},
\label{eq:gHeff}
\end{equation}
where $\Gamma_{H\to X}$ is the decay partial width of $H\to X$\footnote{Each Higgs effective coupling is related to one $\kappa$ parameter in the $\kappa$-formalism \cite{deBlas:2019rxi}.} The electroweak
effective couplings, $g_L^f$ and $g_R^f$ for each fermion $f$, are defined similarly through the partial decay widths of $Z\to f_L\bar{f}_L$ and $Z\to f_R\bar{f}_R$ where $f_L$ and $f_R$ are the left- and right-handed fermion, respectively.

Fit-2 is focused on probing 4-fermion interactions which are present in many BSM models 
with new gauge bosons that couple to the SM fermions. The framework is based on
the study in Ref.~\cite{Falkowski:2017pss} and extends it by including measurements
at future colliders.
It involves both 4-fermion operators
and 2-fermion operators that modify $W/Z$ vertices, around 60 parameters in total. While the vertex part has significant overlap with Fit-1, Fit-2 extends the scope by relaxing the
universality assumption that was imposed in Fit-1 for the first and second generation quarks.
Essentially Fit-2 does not apply any flavor assumption besides focusing only on the diagonal part (\ie\  ignoring flavor violating effects)\footnote{
See Ref.~\cite{Efrati:2015eaa} for an earlier study of a global fit to data with generic flavor structure.}. The effect of 4-fermion operators grows quadratically as 
energy increases, so that this global fit can provide a very good measure of the merit of
running at high energies at future colliders. There are degeneracies which can only be lifted
by low-energy measurements; thus this fit is also a place to study the synergy
between future colliders and low-energy experiments. The renormalization group evolution of the relevant operators at different scales are properly taken into account. The results from Fit-2 are
presented in terms of 1-$\sigma$ bounds directly on the Wilson Coefficients of 4-fermion operators as well as the precision on electroweak effective couplings.
Fit-2 results can also be interpreted in terms of the bounds on $O_{2W}$ 
and $O_{2B}$ operators (defined in~\cite{Giudice:2007fh})\footnote{Using the equations of motion, these operators lead to charged and neutral current four-fermion contact interactions with flavor universality.} which 
correspond to the oblique parameters $W$ and $Y$~\cite{Barbieri:2004qk}.

The top-quark sector has essentially been excluded in the scope of Fit-1 and Fit-2. 
Fit-3 is thus focused on top-quark electroweak couplings and $eett$ 4-fermion operators,
by considering around 20 operators that are directly related to top-quark or the third generation quarks. The top quark has currently not yet been directly produced at any 
lepton colliders, so that top-quark measurements at hadron colliders play an essential role.
Fit-3 also allows one to study the synergy between HL-LHC and future
lepton colliders where top-quark can be directly produced. There have already been
global SMEFT fits to current data including the interplay between the top-quark sector and Higgs/EW sectors~\cite{Ellis:2020unq,Ethier:2021bye}. The interplay
may become more subtle when loop effects from top-quark operators in the Higgs/EW observables are included~\cite{Vryonidou:2018eyv,Durieux:2018ggn,Jung:2020uzh}. 
The Fit-3 for Snowmass 2021 extends these studies by focusing on future colliders, but it is restricted to a more limited set of observables and operators since not all ingredients that are needed for 
such a combined top-quark/Higgs/EW fit are technically ready. Nevertheless, it is useful for studying the interplay of the top-Yukawa coupling with the Higgs/EW sector. 

A few independent sets of codes have been developed to do the global fits, 
using HEPfit~\cite{DeBlas:2019ehy}, Mathematica and C++, each using different statistical models\footnote{The global fit code by HEPfit performs a Bayesian analysis following Markov Chain Monte Carlo procedures with the logarithm of the likelihood function built from measurement projections; the Mathematica code relies on a $\chi^2$ constructed from all observables; and the  C++ code obtains the fitting parameter uncertainties directly from their covariance matrix.}. Cross checks has been performed extensively and excellent consistency has been achieved for the fit results. Since the focus here is on the projected precision and new-physics reach, the central values of all input observables are by default set to the SM expectations. Nevertheless it has been confirmed in Fit-2 that identical uncertainties are obtained if the central values of input observables take their current measurement values from the PDG.

\subsection{Collider scenarios and observables}

The colliders scenarios that are considered in the global fits include the HL-LHC and future $e^+e^-$ colliders
as shown in Tab.~\ref{tab:eecoll}. In addition, future muon colliders are also included in Fit-1 with three 
scenarios: 1 ab$^{-1}$ at 3 TeV; 10 ab$^{-1}$ at 10 TeV; 10 ab$^{-1}$ at 10 TeV plus 20 fb$^{-1}$ at
125 GeV\footnote{Another scenario which combines a 10 TeV muon collider with a future $e^+e^-$ machine is in preparation.}. 
No muon collider scenarios have been considered for Fit-2 and Fit-3, partly due to a lack of suitable studies of input observables\footnote{However, a study of the sensitivity of muon colliders to new 4-fermion interactions was performed in Ref.~\cite{Chen:2022msz} in the framework of the $W/Y$ parameters, which is more constrained than the SMEFT framework.}.
The projected uncertainties of various input observables are mainly supplied by the corresponding 
collider collaborations~\cite{Baer:2013cma,Bambade:2019fyw,ILCInternationalDevelopmentTeam:2022izu,Linssen:2012hp,Charles:2018vfv,Abada:2019zxq,Bernardi:2022hny,CEPCStudyGroup:2018ghi,Cheng:2022zyy,Aime:2022flm,MuonCollider:2022xlm,deBlas:2022aow}. The list of input observables in each fit is too lengthy to be 
included in this report, and we will show here only a few typical examples.
The complete list and details can be found in~\cite{GlobalFitWhitePaper}.

\setcounter{footnote}{0}
The input observables for Fit-1 include: EWPOs as shown in Tab.~\ref{tab:ewpocomp}; Higgs production and decay rates, as shown partially in Tab.~\ref{tab:LHC_HiggsSxBR} for HL-LHC and Tab.~\ref{tab:CC_Higgs240} for FCC-ee and CEPC\footnote{There are many more tables for Higgs inputs at other energies and other colliders \cite{GlobalFitWhitePaper} which are not explicitly listed here for conciseness.}; optimal observables for $e^+e^-\to W^+W^-$~\cite{Diehl:1993br}. The input observables for Fit-2 include: EWPOs as above; cross section and forward-backward asymmetry in $e^+e^-\to f\bar{f}$ off the $Z$-pole\footnote{Due to insufficient inputs from community, a common analysis was performed to obtain those uncertainties for all future $e^+e^-$ which however included only statistical errors.}; low-energy observables as shown in Tab.~\ref{tab:leobs}. Fit-3 includes the following observables: cross section or differential cross section for $t\bar{t}$, single-top, $t\bar{t}Z$ and $t\bar{t}\gamma$ production
at (HL-)\linebreak[0] LHC; cross section and forward-backward asymmetry in $e^+e^-\to b\bar{b}$; optimal observables in $e^+e^-\to t\bar{t}\to bW^+\bar{b}W^-$~\cite{Durieux:2018tev}; cross section for $e^+e^-\to t\bar{t}H$. 

It is important to ensure the consistency among inputs provided by different collider collaborations. One example about common systematic errors in $A_b$ measurements has been elaborated in Section 2.2. Another example is shown in Tab.~\ref{tab:CC_Higgs240}, where we can compare the direct inputs by one collaboration with extrapolated inputs from another collaboration, as illustrated for FCC-ee240 direct inputs and ILC extrapolations (numbers in brackets) in the second column. Even though the $BR_{\gamma Z}$ input was missing in the FCC-ee documents, the extrapolated projection is included in the global fit since this observable turns out to play a sensitive role. More examples about the procedures that were taken to ensure consistency on inputs can be found in Ref.~\cite{GlobalFitWhitePaper}.

\begin{table}[p]
  \centering
  \begin{tabular}{|c|c|c|c|c|c|}
    \hline
    HL-LHC    &   \multicolumn{5}{|c|}{3 ab$^{-1}$ ~ ATLAS+CMS} \\
    \hline
    Prod.     &    ggH     &   VBF   &    WH   &    ZH   & ttH \\
    $\sigma$  & - & -  & - & - & - \\
    $\sigma\times BR_{bb}$  &   19.1  &  -  & 8.3  & 4.6  & 10.2 \\
    $\sigma\times BR_{cc}$  &   -     &  -  & -    & -    & - \\   
    $\sigma\times BR_{gg}$  &   -     &  -  & -    & -    & - \\     
    $\sigma\times BR_{ZZ}$  &   2.5   &  9.5  & 32.1  & 58.3  & 15.2 \\
    $\sigma\times BR_{WW}$  &   2.5   &  5.5  & 9.9   & 12.8  & 6.6 \\    
    $\sigma\times BR_{\tau\tau}$  &   4.5   &  3.9  & -  & -  & 10.2 \\    
    $\sigma\times BR_{\gamma\gamma}$  &   2.5   &  7.9  & 9.9  & 13.2  & 5.9 \\    
    $\sigma\times BR_{\gamma Z}$  &   24.4   &  51.2  & -  & -  & - \\    
    $\sigma\times BR_{\mu\mu}$  &   11.1   &  30.7  & -  & -  & - \\    
    $\sigma\times BR_{inv.}$  &   -   &  2.5  & -  & -  & - \\ 
    $\Delta \mh$  & 30 MeV & - & - & - & - \\
    \hline    
  \end{tabular}
\caption{Projected uncertainties of Higgs observables at HL-LHC for the leading five production channels and various decay modes. Numbers are in \%, except for $\mh$.}
\label{tab:LHC_HiggsSxBR}
\end{table}

\begin{table}[p]
  \centering
  \begin{tabular}{|c|c|c|c|c|}
    \hline
        &   \multicolumn{2}{c|}{FCCee240  5ab$^{-1}$} & \multicolumn{2}{c|}{CEPC240  20ab$^{-1}$}\\
    \hline
    Prod.     &    $ZH$     &   $\nu\nu H$   &    $ZH$   &    $\nu\nu H$ \\
    $\sigma$  & 0.5(0.537) & -  & 0.26 & - \\
    $\sigma\times BR_{bb}$  &   0.3(0.380)  &  3.1(2.78)  & 0.14  & 1.59 \\
    $\sigma\times BR_{cc}$  &   2.2(2.08)     &  -  & 2.02    & -    \\   
    $\sigma\times BR_{gg}$  &   1.9(1.75)     &  -  & 0.81    & -    \\     
    $\sigma\times BR_{ZZ}$  &   4.4(4.49)   &  -  & 4.17  & -  \\
    $\sigma\times BR_{WW}$  &   1.2(1.16)   &  -  & 0.53   & -  \\    
    $\sigma\times BR_{\tau\tau}$  &   0.9(0.822)   &  -  & 0.42  & - \\    
    $\sigma\times BR_{\gamma\gamma}$  &   9(8.47)   &  -  & 3.02  & - \\    
    $\sigma\times BR_{\gamma Z}$  &   (17$^*$) & -  & 8.5  & - \\    
    $\sigma\times BR_{\mu\mu}$  &   19(17.9)   &  -  & 6.36  & - \\    
    $\sigma\times BR_{inv.}$  &  0.3(0.226)   &  -  & 0.07  & - \\ 
%    $\Delta m_H$  & 10-20 MeV & - & - & - & - \\    
 \hline    
  \end{tabular}
\caption{Projected uncertainties of Higgs observables at FCCee240 and CEPC240 in the two leading production channels and various decay modes. Numbers are in \%. The numbers in brackets are extrapolated from the projections at ILC250.}
\label{tab:CC_Higgs240}
\end{table}

\begin{table}[t]
\begin{center}
\begin{adjustbox}{max width = \textwidth}
\renewcommand{\arraystretch}{1.5}
\begin{tabular}{|c|c|c|c|c|}
\hline
{Process} & {Observable} & {Experimental value}   &   {Ref.}   &  {SM prediction}
 \\  \hline  \hline
\multirow{2}{*}{$\stackrel{(-)}{\nu}_\mu-e^-$ scattering} & $g_{LV}^{\nu_\mu e}$ & $-0.035\pm0.017$ &\multirow{2}{*}{CHARM-II\cite{CHARM-II:1994dzw}}    &  $-0.0396$\cite{Erler:2013xha}\\
 & $g_{LA}^{\nu_\mu e}$ & $-0.503\pm0.017$  &   &  $-0.5064$\cite{Erler:2013xha}\\\hline
\multirow{2}{*}{$\tau$ decay} & $\frac{G_{\tau e}^2}{G_F^2}$ & $1.0029\pm0.0046$ & \multirow{2}{*}{PDG2014\cite{ParticleDataGroup:2014cgo}}    &  \multirow{2}{*}{$1$}\\ %PDG
 & $\frac{G_{\tau \mu}^2}{G_F^2}$ & $0.981\pm0.018$  &  & \\\hline
\multirow{6}{*}{Neutrino scattering} & $R_{\nu_\mu}$ & $0.3093\pm0.0031$ & \multirow{2}{*}{CHARM ($r=0.456$)\cite{CHARM:1987pwr}} & 0.3156\cite{CHARM:1987pwr} \\
 & $R_{\bar{\nu}_\mu}$ & $0.390\pm0.014$ & & 0.370\cite{CHARM:1987pwr} \\
 \cline{2-5}
 & $R_{\nu_\mu}$ & $0.3072\pm0.0033$ & \multirow{2}{*}{CDHS ($r=0.393$)\cite{Blondel:1989ev}}  & 0.3091\cite{Blondel:1989ev}\\
 & $R_{\bar{\nu}_\mu}$ & $0.382\pm0.016$ & & 0.380\cite{Blondel:1989ev}\\
 \cline{2-5}
  & $\kappa$ & $0.5820\pm0.0041$ & CCFR\cite{CCFR:1997zzq} & 0.5830\cite{CCFR:1997zzq}\\
 \cline{2-5}
  & $R_{\nu_e\bar{\nu}_e}$ & $0.406^{+0.145}_{-0.135}$ & CHARM\cite{CHARM:1986vuz} & 0.33\cite{ParticleDataGroup:2016lqr}\\
\hline
 \multirow{9}{*}{\parbox{8em}{Parity-violating scattering}} & $(s^2_w)^\text{M\o ller}$ & $0.2397\pm0.0013$ &SLAC-E158\cite{SLACE158:2005uay}    &  $0.2381\pm0.0006$\cite{Czarnecki:1995fw}\\
 \cline{2-5}
 & $Q_W^{\rm Cs}(55,78)$ & $-72.62\pm0.43$ & PDG2016\cite{ParticleDataGroup:2016lqr} & $-73.25\pm0.02$\cite{ParticleDataGroup:2016lqr} \\
 \cline{2-5}
 & $Q_W^{\rm p}(1,0)$ & $0.064\pm0.012$ & QWEAK\cite{Qweak:2013zxf} & $0.0708\pm0.0003$\cite{ParticleDataGroup:2016lqr} \\
 \cline{2-5}
 & $A_1$ & $(-91.1\pm4.3)\times10^{-6}$ & \multirow{2}{*}{PVDIS\cite{PVDIS:2014cmd}} & $(-87.7\pm0.7)\times10^{-6}$\cite{PVDIS:2014cmd} \\
 & $A_2$ & $(-160.8\pm7.1)\times10^{-6}$ & & $(-158.9\pm1.0)\times10^{-6}$\cite{PVDIS:2014cmd}\\
 \cline{2-5}
 & \multirow{2}{*}{$g_{VA}^{eu}-g_{VA}^{ed}$} & $-0.042\pm0.057$ & SAMPLE ($\sqrt{Q^2}=200$\,MeV)\cite{Beise:2004py} & -0.0360\cite{ParticleDataGroup:2016lqr}\\
 & & $-0.12\pm0.074$ & SAMPLE ($\sqrt{Q^2}=125$\,MeV)\cite{Beise:2004py} & 0.0265\cite{ParticleDataGroup:2016lqr}\\
 \cline{2-5}
 & \multirow{2}{*}{$b_{\rm SPS}$} & $-(1.47\pm0.42)\times10^{-4}\rm\,GeV^{-2}$ & SPS $(\lambda=0.81)$\cite{Argento:1982tq} & $-1.56\times10^{-4}\rm\,GeV^{-2}$\cite{Argento:1982tq} \\
 & & $-(1.74\pm0.81)\times10^{-4}\rm\,GeV^{-2}$ & SPS $(\lambda=0.66)$\cite{Argento:1982tq} & $-1.57\times10^{-4}\rm\,GeV^{-2}$\cite{Argento:1982tq} \\
\hline
\multirow{2}{*}{$\tau$ polarization} & $\mathcal{P}_\tau$ & $0.012\pm0.058$ & \multirow{2}{*}{VENUS\cite{VENUS:1997cjg}} & 0.028\cite{VENUS:1997cjg} \\
  & $\mathcal{A}_{\mathcal{P}}$ & $0.029\pm0.057$ &  & 0.021\cite{VENUS:1997cjg} \\
\hline
\rule{0mm}{1.8em}\parbox{8em}{Neutrino trident\newline production} & \parbox{6em}{\raggedleft $\frac{\sigma}{\sigma^{\rm SM}}(\nu_\mu \gamma^*\to\nu_\mu\mu^+\mu^-)$} & $0.82\pm0.28$ & CCFR\cite{CHARM-II:1990dvf,CCFR:1991lpl,Altmannshofer:2014pba} & 1 \\
\hline
%\multirow{2}{*}{$d_j\to u_i\ell\bar{\nu}_\ell$} & $\tilde{V}_{ud}$ & \cite{Gonzalez-Alonso:2016etj} & \cite{Gonzalez-Alonso:2016etj} & 0.974353 \\
{$d_I\to u_J\ell\bar{\nu}_\ell(\gamma)$}& $\epsilon_{L,R,S,P,T}^{de_J}$ & See Ref.~\cite{GlobalFitWhitePaper} & \cite{Gonzalez-Alonso:2016etj} & 0 \\
\hline
%\multirow{2}{*}{Neutrino scattering} & & & & \\
\end{tabular}
\end{adjustbox}
\end{center}
\caption{Low-energy observables included in Fit-2.}
\label{tab:leobs}
\end{table}

\subsection{Results}

\begin{figure}[p]
    \includegraphics[width=\textwidth]{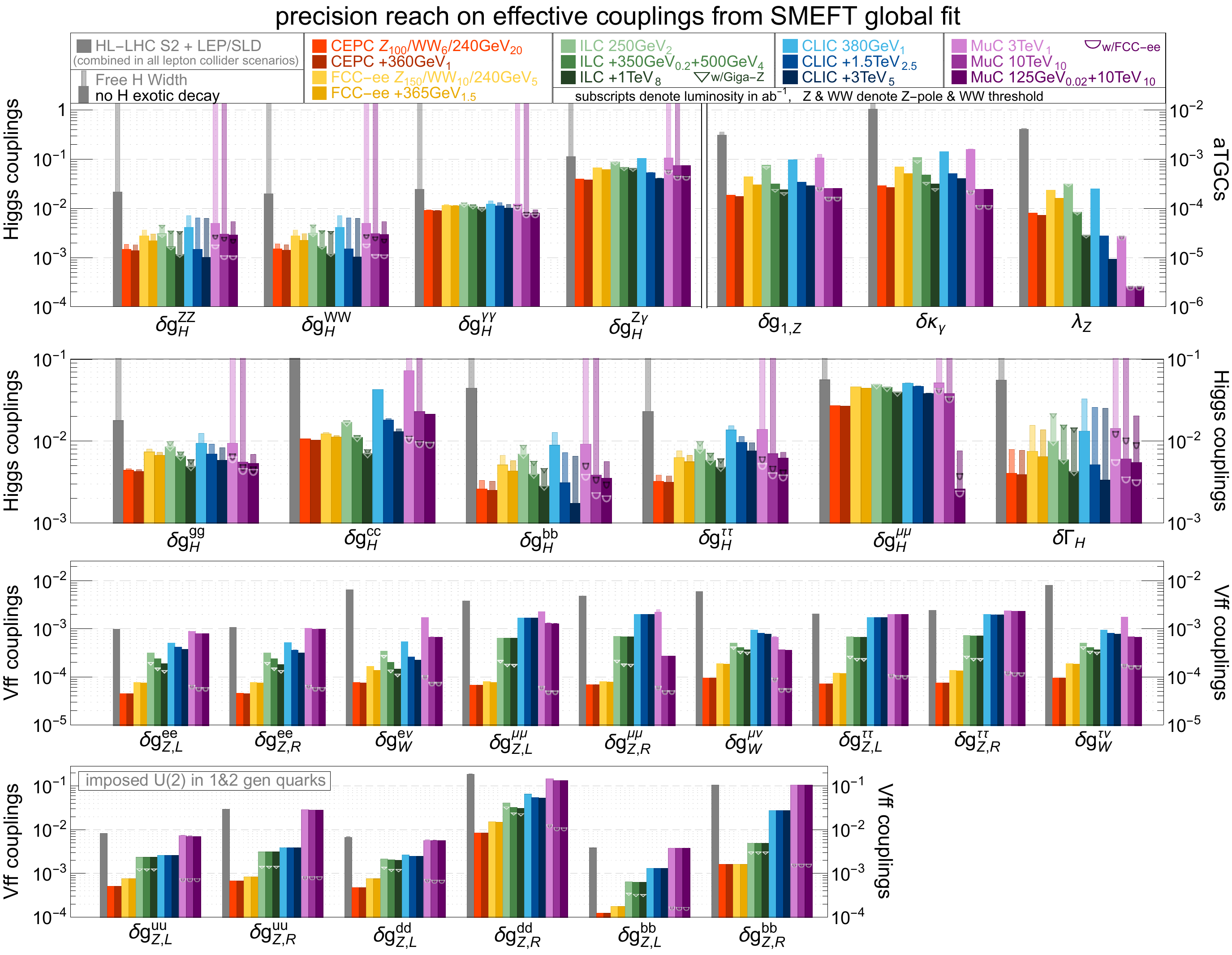}
    \caption{Precision reach on Higgs and electroweak effective couplings from a SMEFT global analysis of the Higgs and EW measurements at various future colliders.  The wide (narrow) bars correspond to the results  
    from the constrained-$\Gamma_H$ (free-$\Gamma_H$) fit. The HL-LHC and LEP/SLD measurements are combined with all future lepton collider scenarios.  For $e^+e^-$ colliders, the high-energy runs are always combined with the low energy ones.  For the ILC, the (upper edge of the) triangle mark shows the results for which a Giga-Z run is also included.  For the muon collider, three separate scenarios are considered.  The subscripts in the collider scenarios denote the corresponding integrated luminosity of the run in ${\rm ab}^{-1}$. Note the Higgs total width measurement from the off-shell Higgs processes at the HL-LHC is not included in the global fit.}
    \label{fig:Fit1-HiggsEW}
\end{figure}

\begin{figure}[tb]
    \includegraphics[width=\textwidth]{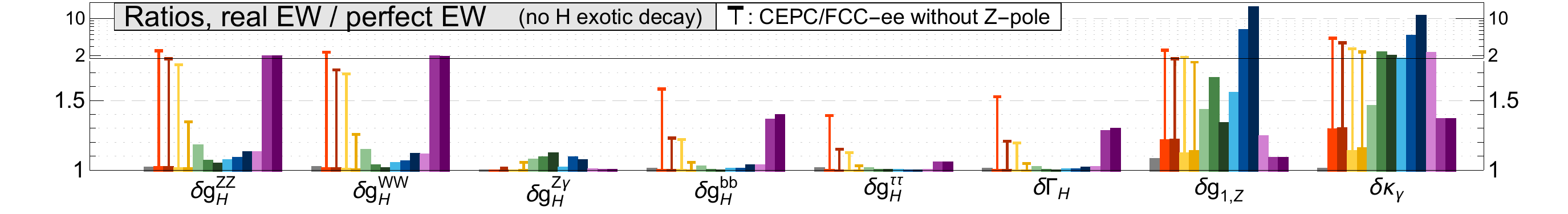}%
    \vspace{-1ex}
    \caption{Ratios of the measurement precision (shown in \autoref{fig:Fit1-HiggsEW}) to the one assuming perfect EW measurements (Z pole + W mass/width) in the constrained-$\Gamma_H$ fit.  Results are only shown for Higgs couplings and aTGCs with ratios significantly larger than one.  For CEPC/FCC-ee, we also show (with the thin bars) the results without their Z-pole measurements.}
    \label{fig:Fit1-PZratio}
\end{figure}

The results of Fit-1 are shown in Fig.~\ref{fig:Fit1-HiggsEW} for electroweak and Higgs effective couplings as defined in Sec. 4.1.
They are plotted as 1$\sigma$ relative uncertainties for two cases of
global fits: the wider (narrower) bars assume that
the Higgs total width is constrained (free)\footnote{Allowing the Higgs total width to be a free parameter accounts for the possibility of non-standard Higgs decays into BSM particles.}. The grey bars represent
the expectation from HL-LHC measurements while colored bars are for 
various future lepton colliders as indicated in the legend. 
The input measurements from HL-LHC are always included in the fits for future colliders. For each future collider, results are shown for various running scenarios with measurements in earlier stages always combined with the later ones (denoted by the ``+'' symbol in the legend), except for the muon colliders. 
The electroweak effective couplings include: $Z$ couplings to left- and
right-handed leptons (for $e,\mu,\tau$) and quarks (for $u,d,b$\footnote{Note that the universality assumption for $1^{\rm st}$ and $2^{\rm nd}$ generation quarks implies that couplings for $c$($s$) quarks are as same as for $u$($d$) quarks}); $W$ couplings to leptons\footnote{$Z$ couplings to neutrinos and $W$ couplings to quarks are not listed separately since they are related to the other couplings for operators up to dimension 6 in SMEFT.}. 
The Higgs effective couplings include Higgs couplings to $ZZ,WW,\gamma\gamma,Z\gamma,gg,cc,bb,\tau\tau,\mu\mu$ as well as
the Higgs total width. $Z$ couplings to top quarks and the top-Yukawa coupling will be discussed in Fit-3 results. In addition to the above effective couplings, which are independent of the operator basis, the figure also shows three anomalous triple gauge couplings (aTGCs): $g_{1,Z},\kappa_\gamma,\lambda_Z$. The exact definition of each parameter or Wilson coefficient shown in this and following plots can be found in Ref.~\cite{GlobalFitWhitePaper}. 

From the Fit-1 results one can deduce that future $e^+e^-$ colliders can improve our knowledge of electroweak effective couplings by a few orders of magnitude. The improvement will mainly come from dedicated runs at the $Z$-pole and $WW$-threshold for circular $e^+e^-$, and lower-energy stages at linear $e^+e^-$ via the radiative return process. The higher energy stages of any $e^+e^-$ have little impact on most of the effective couplings except for: $Z$ couplings to electrons, which as illustrated in Fig.~\ref{fig:EFT_OP_cHE} are related to $eeZH$ contact interactions which increase quadratically with energy; $W$ couplings to leptons, simply due to increased statistics from $WW$ production. At muon colliders, the expected improvements will be mainly for $Z$ couplings to muons and $W$ couplings to leptons for the same reason as for higher-energy stages of $e^+e^-$. The aTGCs can benefit a lot from higher-energy stages of $e^+e^-$ or muon colliders, in particular for $\lambda_Z$, which is sensitive to the transverse modes of $W$ bosons. In general, circular $e^+e^-$ can deliver the best precision for electroweak effective couplings, while linear $e^+e^-$ can bring comparable improvements in particular when the Giga-Z option with beam polarization is included.

For results on Higgs effective couplings, HL-LHC will push the constraints to 2-5\% for many couplings while future $e^+e^-$ or muon colliders will improve further to 1\% or below. In addition, future $e^+e^-$ can bring a qualitative difference when the Higgs total width is a free parameter in the global fit\footnote{Alternatively speaking, the Higgs total width can be indirectly determined at future $e^+e^-$ without any assumption on possible decay modes.}. High-energy muon colliders can also bring this advantage when a dedicated scan at the Higgs pole is included. 
There is a potential at the HL-LHC to determine the Higgs total width using off-shell Higgs measurements~\cite{Caola:2013yja,Campbell:2013una} with an uncertainty of 0.75 MeV~\cite{ATLAS:2022hsp,CMS:2022ley}\footnote{This uncertainty is likely to be improved once the $WW$ channel is employed in addition to the current $ZZ$ analyses.}. This piece of input has not been included in the global fit since the full EFT treatment for this measurement is not yet available \cite{Azatov:2022kbs}. 

It is worth noting the interplay between Higgs couplings and EWPOs as shown in Fig.~\ref{fig:Fit1-PZratio}: For circurlar $e^+e^-$ the achievable precision for Higgs couplings and aTGCs improves by a factor of around 2 when including measurements at the $Z$-pole and $WW$ threshold. The possibility for similar improvements at muon colliders depends on their ability to meausure EWPOs with high precision, which likely would require a dedicated $Z$-pole run.
At linear $e^+e^-$ colliders, the availability of beam polarization helps to break degeneracies in the SMEFT parameter space already from measurements at 250/380~GeV alone, and thus the impact of EW precision data on the other couplings is less significant. It is also worth to note the synergies on Higgs rare decays ($H\to\gamma\gamma,Z\gamma,\mu\mu$) between HL-LHC and future lepton colliders which play an important role in the global fits.

\begin{figure}%[t]
    \includegraphics[width=\textwidth]{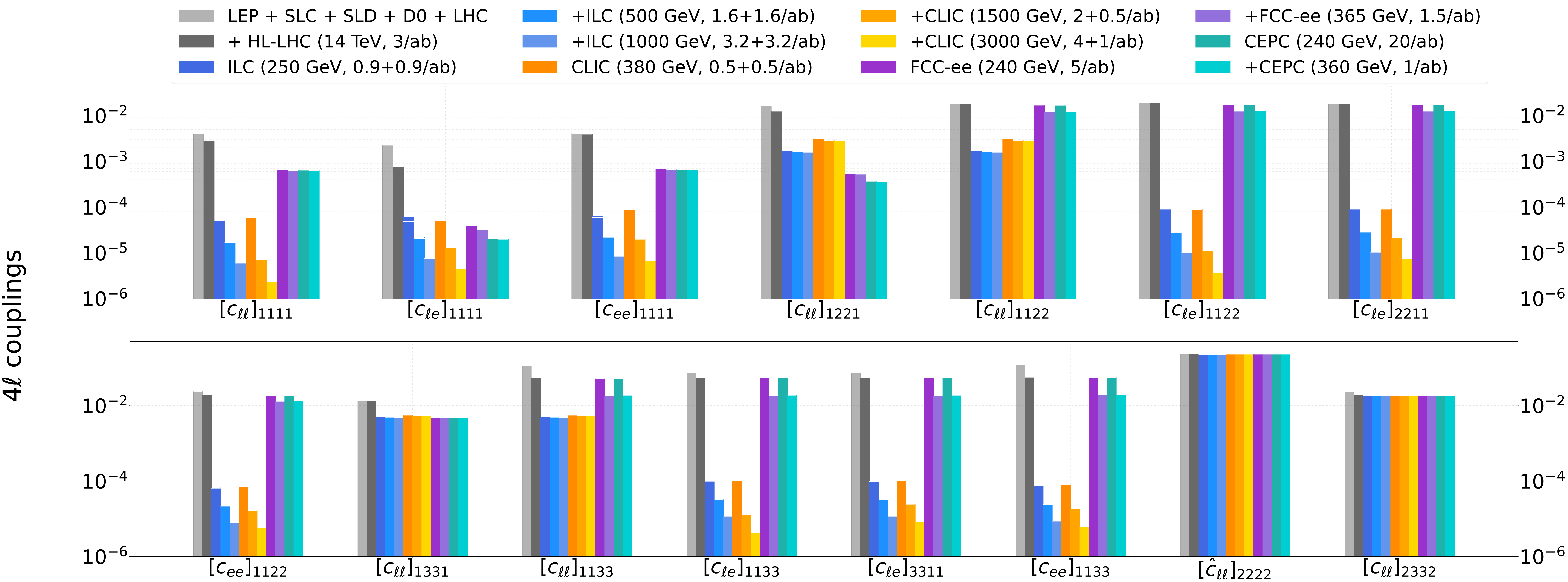}
    \caption{Precision reach on the 4-fermion operators and electroweak effective couplings from a SMEFT global fit at various future lepton colliders. "LEP+SLC+SLD" represents current measurements which are always combined in the future collider scenarios. The horizontal white line for ILC illustrates the global fit results when the pole observables from its Giga-Z option are included.}
    \label{fig:Fit2-4f1}
\end{figure}

The results of Fit-2 are shown in Fig.~\ref{fig:Fit2-4f1}--\ref{fig:Fit2-4f3} for present measurement results and future $e^+e^-$ projections including the following: $1\sigma$ relative uncertainties for electroweak effective couplings\footnote{With the flavor assumption relaxed, $Z$ couplings to $u,d,c,s$ are treated separately.}; $1\sigma$ absolute uncertainties for the Wilson Coefficients of 4-fermion operators. The conclusions for electroweak couplings are consistent with that from Fit-1. For 4-fermion operators the higher-energy stages of future $e^+e^-$ will bring much more profound improvements since the sensitivity to 4-fermion interactions grows quadratically as energy increases. Even with the same energy, linear $e^+e^-$ can deliver much higher sensitivity than circular ones since there are crucial degenericies among 4-fermion operators that the beam polarization can help lift. 

The results of Fit-3 are shown in Fig.~\ref{fig:Fit3-top1}--\ref{fig:Fit3-top3} for 95\% C.L. bounds on the Wilson Coefficients of various top-quark operators, for LHC and future $e^+e^-$. LHC will bring invaluable constraints on many top-quark operators while those related to top-quark electroweak couplings will be improved by future $e^+e^-$ in particular when the collision energies above 500 GeV are envisaged. There are a range of $eett$ 4-fermion operators whose degeneracties can not be lifted unless there are measurements with at least two distinct energies well above the $t\bar{t}$ threshold. The uncertainty of top-Yukawa coupling is encapsulated in the bound on $C_{t\phi}$, for which synergies between LHC and future $e^+e^-$ play an important role. There will be no direct constraint on the top Yukawa coupling from circular $e^+e^-$ below 500 GeV; thus the projection from HL-LHC will provide the best constraints in such a scenario. The converted $1\sigma$ relative uncertainties on the top Yukawa coupling from the global and indivual fits are shown in Tab.~\ref{tab:Fit3-ttH}. It is worth noting that ILC running at 550 GeV would improve top-Yukawa coupling significantly compared to running at nominal 500 GeV. 

The result on the triple Higgs coupling ($\lambda_{hhh}$) from a global fit performed by the ESG can be found in Fig.~11 of Ref.~\cite{deBlas:2019rxi}. A fit of $\lambda_{hhh}$ has currently not been included in Fit-1, but is not expected to be much different. 

Theory uncertainties may play a significant role in either electroweak couplings or Higgs couplings in the global fits. 
%A complete update is not yet available. 
The impact on Higgs couplings was discussed by the ESG in Tab.~10 and 11 in Ref.~\cite{deBlas:2019rxi}. One major challenge will come from the intrinsic error of the SM prediction for $e^+e^-\to ZH$ cross section, which would be around 0.5\% with NNLO EW correction and will be significant enough to affect the precision of $HZZ$ and $HWW$ couplings. In addition, an important parametric error is due to the bottom-quark mass uncertainty ($\sim$13 MeV) which would affect the bottom-Yukawa coupling precision. The Higgs mass uncertainty is another source of parametric error, which will become subdominant if a precision of about 10~MeV can be reached. 
The detailed impact of theory errors on various effective 
Higgs and EW couplings is quantified in the context of Fit-1 and summarized in Fig.~\ref{fig:theory-error}. The parametric errors for all EWPOs and
Higgs production and decay rates are included, in particular the projected uncertainties of the bottom-quark and Higgs masses mentioned above, the strong coupling constant ($\Delta\alphas=0.0002$), charm quark mass ($\Delta m_c=7$ MeV) and top quark mass ($\Delta \mt=50$ MeV). The intrinsic theory errors for EWPOs are as introduced in Tab.~\ref{tab:ewpoth}, while the intrinsic errors for Higgs production and decay rates can be found in Ref.~\cite{deBlas:2019rxi}. It is worth noting that only the theory errors for SM predictions are included while the theory errors related to SMEFT (\eg\ the uncertainty due to truncation up to linear effects from dimension-6 operators) are not evaluated here. More details can be found in Ref.~\cite{GlobalFitWhitePaper}. 
We can clearly see the need to improve in particular the intrinsic theory errors in order to make use of the precision that future experiments could deliver.

The results from global fits can be also interpreted in terms of constraints on simple BSM benchmark models with a small set of parameters (\ie\ only a small set of SMEFT operators are generated in each model).
Three examples are studied for the Fit-2 global fit results.
The first example considers a flavor-universal 4-fermion contact interaction, which can be described by the $O_{2B}$ operator mentioned in section~\ref{subsec:eftframeworkscope}. Figure~\ref{fig:model-4f} shows the bounds on the scale of this operator one can draw for different future colliders. 
The same bound on $O_{2B}$ can also be interpreted in the $Y$-universal $Z^\prime$
model~\cite{Appelquist:2002mw} as a bound on the gauge coupling $g_{Z^\prime}$ versus $Z^\prime$ mass,
as shown in Fig.~\ref{fig:model-zprime}. The third benchmark model extends the SM by two leptoquark multiplets, one being an SU(2) singlet, while the other is a SU(2) triplet~\cite{Gherardi:2020det,Aebischer:2021uvt}. This model can generate various 2-lepton--2-quark contact interactions by integrating out the heavy leptoquark fields in the $t$-channel. The bounds on the ratios of Yukawa couplings $\lambda_i$ over leptoquark mass $M_i$ are shown in Fig.~\ref{fig:model-leptoquark}, where $i=1$ (3) refers to the singlet (triplet) leptoquark.

%%%JT: 4f results reorganized as 4-lepton, 2-lepton 2-quark and vertex part%%% 

\begin{figure}%[t]
    \includegraphics[width=\textwidth]{plot_Fit2-4f-2l2q.pdf}
    \caption{\autoref{fig:Fit2-4f1} continued.}
    \label{fig:Fit2-4f2}
\end{figure}

\begin{figure}%[t]
    \includegraphics[width=\textwidth]{plot_Fit2-4f-vff.pdf}
    \caption{\autoref{fig:Fit2-4f2} continued.}
    \label{fig:Fit2-4f3}
\end{figure}

\clearpage

\begin{figure}[h!]
%\hspace*{-1.5 cm}%
\includegraphics[trim=100 50 120 100,clip,width=\textwidth]{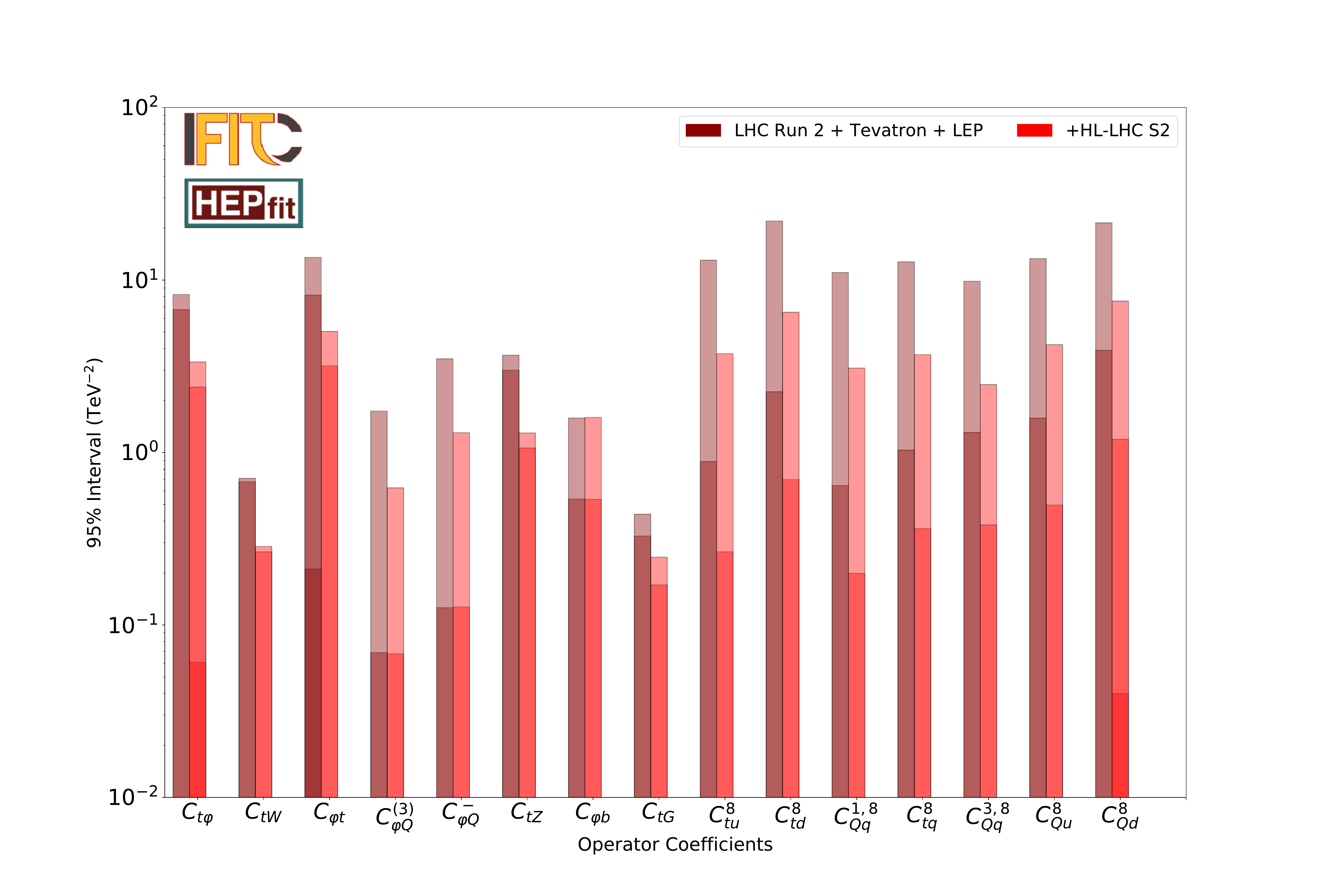}
\caption{The 95\% probability bounds on the Wilson coefficients for dimension-six operators that affect the top-quark production and decay measurements after Run 2 of the LHC (in dark red) and prospects for the bounds expected after completion of the complete LHC program, including the high-luminosity stage (in light red). The individual bounds obtained from a single-parameter fit are shown as solid bars, while the global or marginalised bounds obtained fitting all Wilson coefficients at once are indicated by the full bars (pale shaded region in each bar). }
\label{fig:Fit3-top1}
\end{figure}

\begin{figure}[h!]
\includegraphics[trim=100 50 120 100,clip,width=\textwidth]{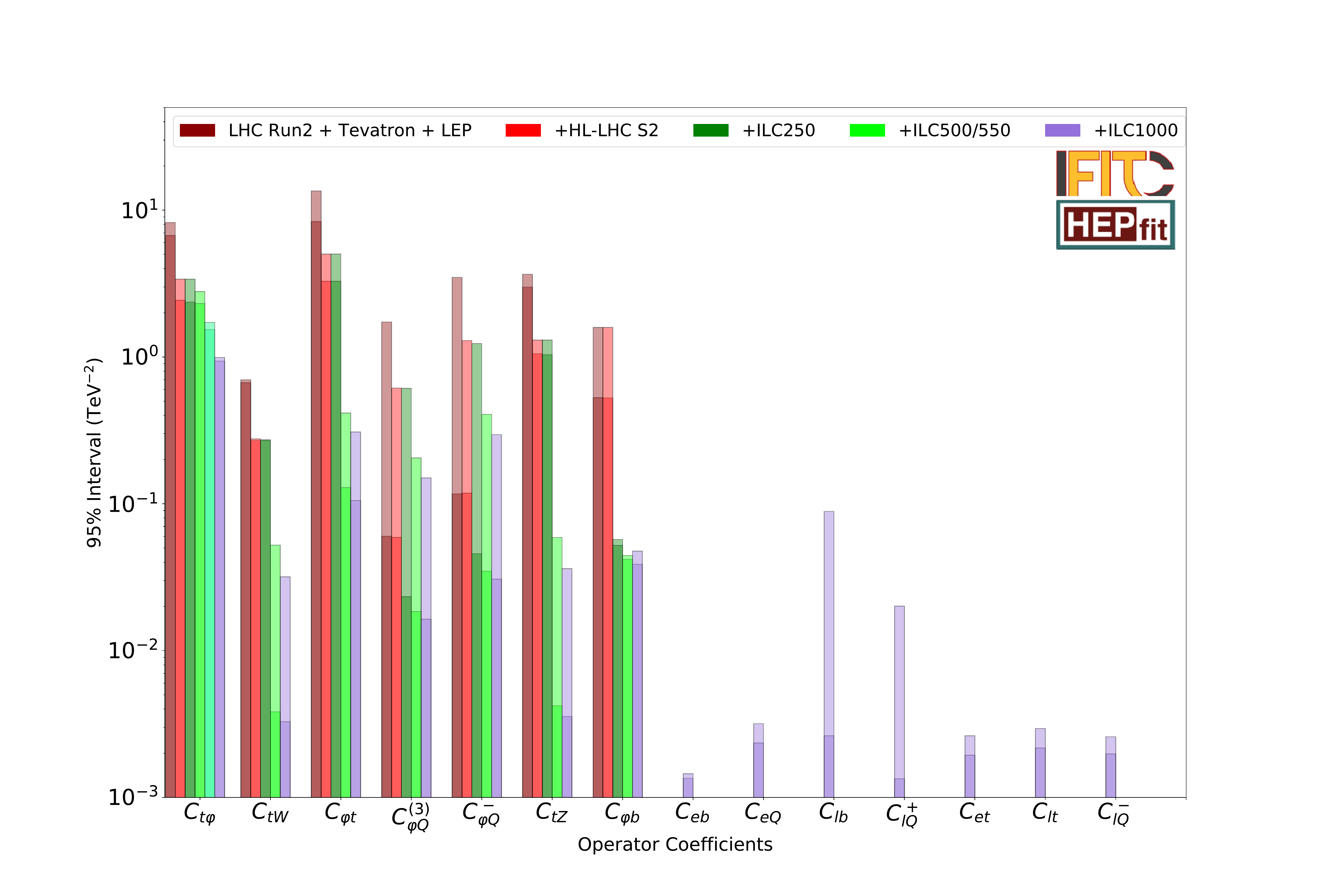}
\caption{Comparison of current LHC constraints on various top-sector Wilson coefficients with HL-LHC ones, and those derived from ILC runs at 250, 500 and 1000~GeV.
The limits on the $q\bar{q}t\bar{t}$ and $C_{tG}$ coefficients are not shown, since the $e^+e^-$ collider measurements considered are not sensitive to them, but all operators are included in the global fit.
The improvement expected from HL-LHC on these coefficients is shown in Fig.~\ref{fig:Fit3-top1}. The additional bar included for $C_{t\varphi}$ in light green shows the effect on this operator of ILC working at 550 GeV. The solid bars provide the individual limits of the single-parameter fit and the shaded ones the marginalised limits of the global fit.}
\label{fig:Fit3-top2}
\end{figure}

\begin{figure}[t!]
%\hspace*{-1.5 cm}%
\includegraphics[trim=100 50 120 150,clip,width=\textwidth]{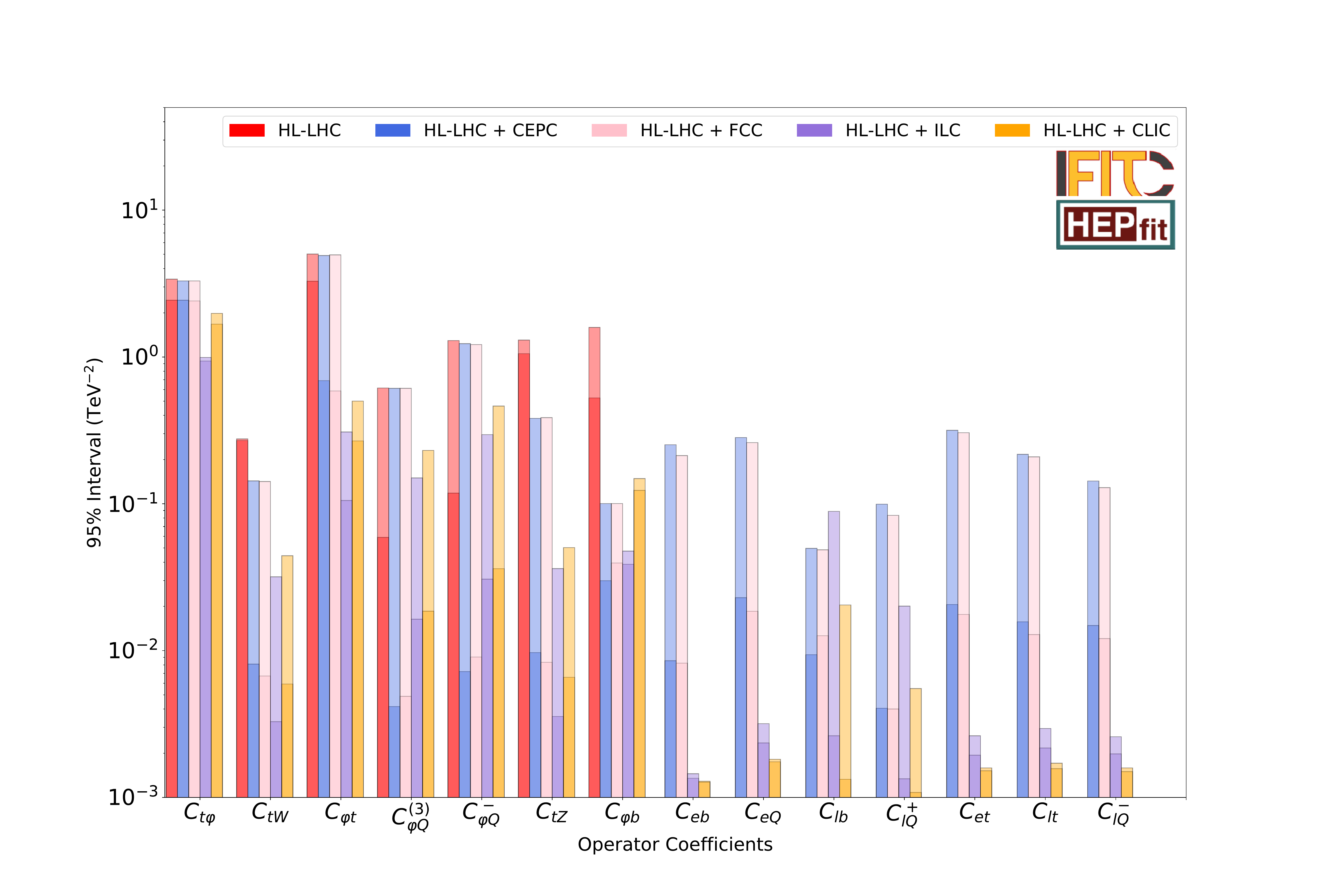}
\caption{Constraints expected on top-sector Wilson coefficients from a combination of HL-LHC and lepton collider data.
The limits on the $q\bar{q}t\bar{t}$ and $C_{tG}$ coefficients are not shown, since the $e^+e^-$ collider measurements considered are not sensitive to them, but all operators are included in the global fit.
The improvement expected from HL-LHC on these coefficients is shown in Fig.~\ref{fig:Fit3-top1}. The solid bars provide the individual limits of the single-parameter fit and the pale shaded ones the marginalised limits of the global fit. The results for ILC and CLIC are based on a combination of both low- and high-energy run scenarios.
}
\label{fig:Fit3-top3}
\end{figure}

\begin{table}[]
    \centering
    \begin{tabular}{c|c|c|c|c|c|c|c|}
    \multicolumn{2}{c|}{ Values in \% units }   & LHC  & HL-LHC & ILC500 & ILC550 & ILC1000 & CLIC  \\\hline
    \multirow{2}{*}{$\delta y_t$} & Global fit  & 12.2 & 5.06   & 3.14   & 2.60   & 1.48   & 2.96  \\
                                  & Indiv. fit  & 10.2 & 3.70   & 2.82   & 2.34   & 1.41   & 2.52
    \end{tabular}
    \caption{
    Uncertainties for the top-quark Yukawa coupling at 68\% probability for different scenarios, in percentage. The ILC500, ILC550 and CLIC scenarios also include the HL-LHC. The ILC1000 scenario includes also ILC500 and HL-LHC.
    }
    \label{tab:Fit3-ttH}
\end{table}

\begin{figure}%[t]
    \includegraphics[width=\textwidth]{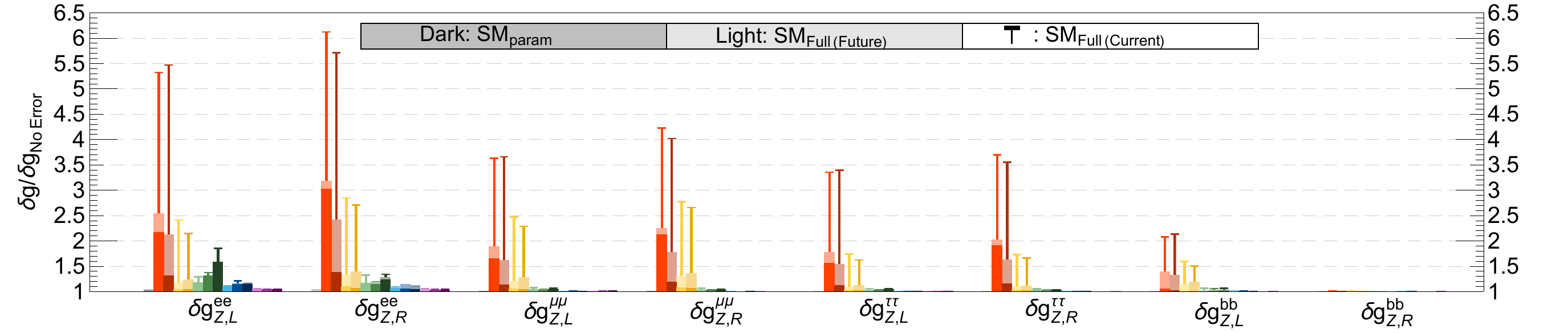} \\
    \includegraphics[width=\textwidth]{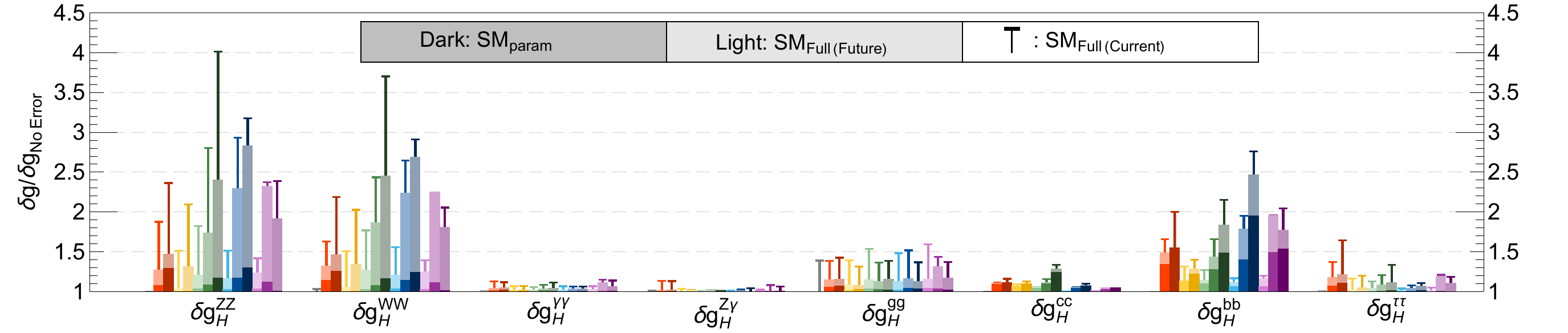}
    \caption{The impact of parametric and intrinsic theory errors on effective EW and Higgs couplings. The impact is plotted in terms of the ratio of uncertainties when theory errors are included in the global fit over the case when they are not included. The dark bars are when only parametric errors are included; the light bars are when both parametric and intrinsic errors as projected in future are included; the thin lines are when current full theory errors are included.}
    \label{fig:theory-error}
\end{figure}

\begin{figure}[p]
\centering
\includegraphics[trim=0 0 0 20,width=.9\textwidth]{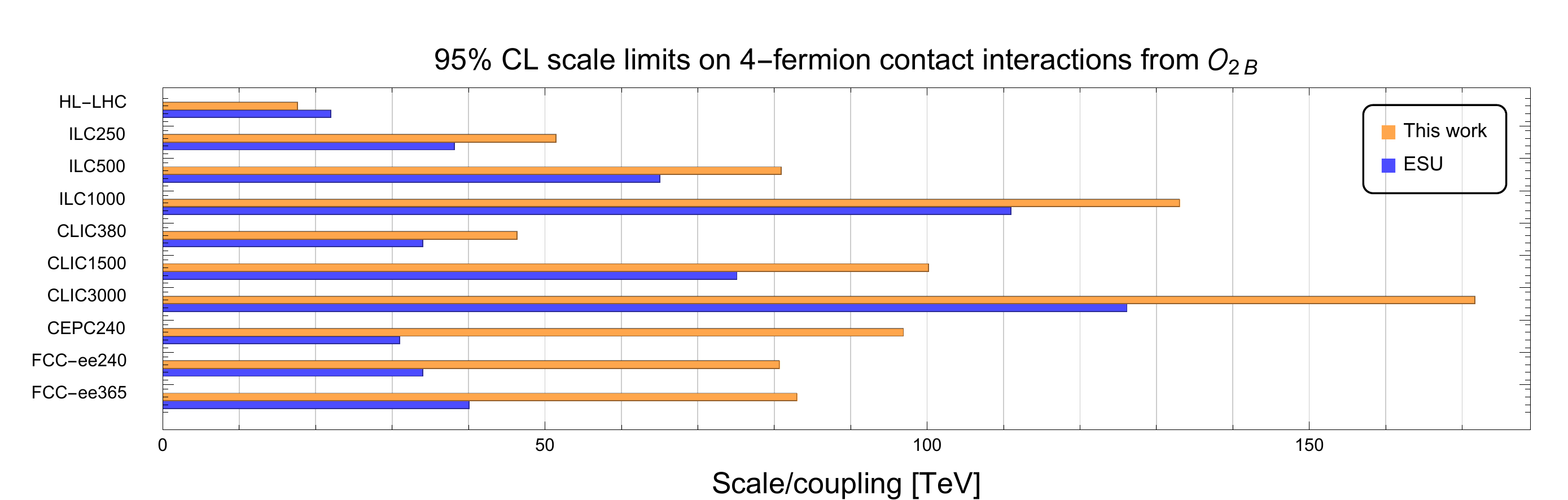}%
\vspace{-2ex}
\caption{95\% C.L. exclusion reach of different colliders on four-fermion contact interactions from the operator $O_{2B}$ (numbers for ESG are taken from Ref.~\cite{EuropeanStrategyforParticlePhysicsPreparatoryGroup:2019qin}).}
\label{fig:model-4f}
\end{figure}

\begin{figure}[p]
\centering
\includegraphics[trim=0 0 0 20,width=.7\textwidth]{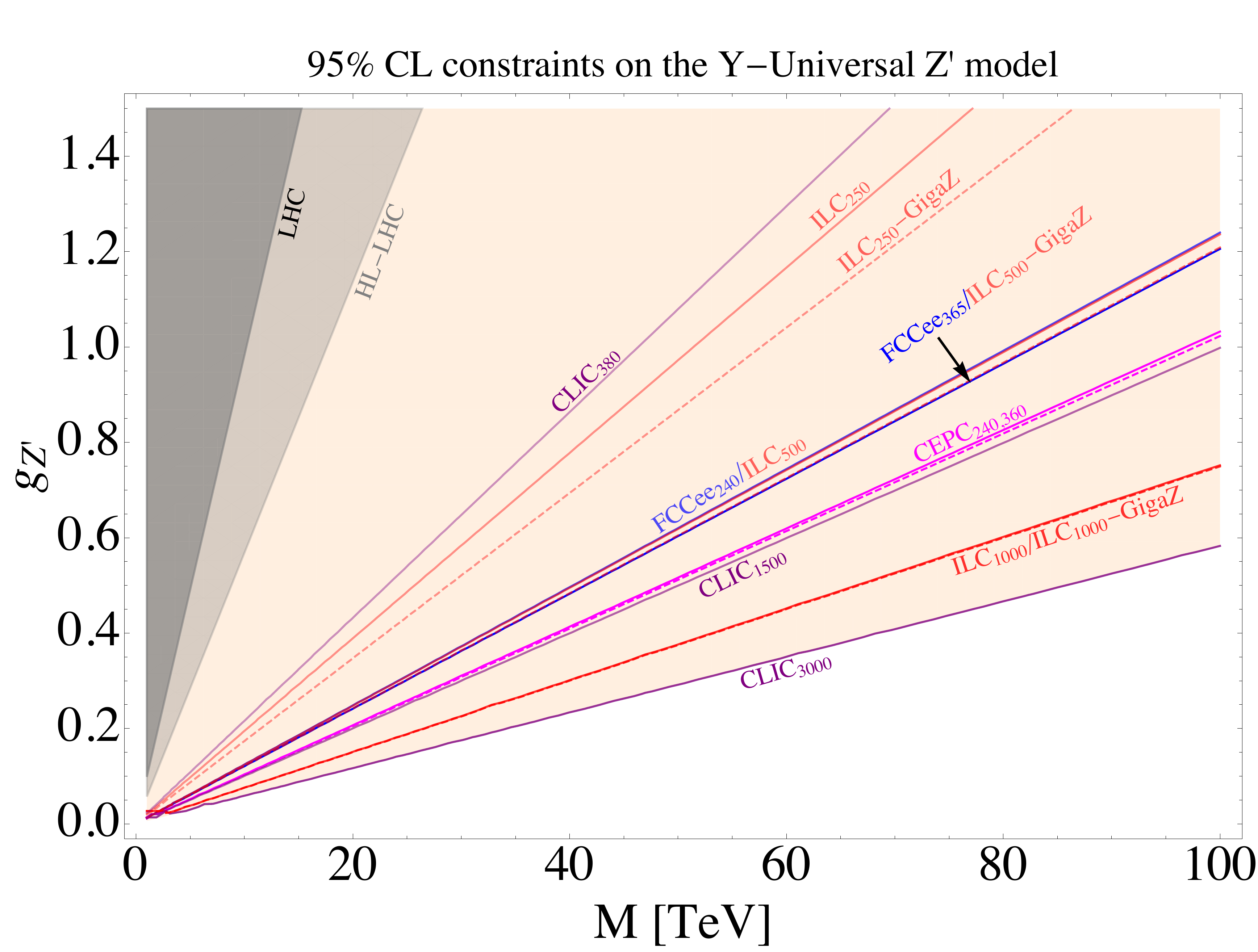}%
\vspace{-1ex}
\caption{95\% C.L. exclusion reach of different colliders on the $Y$-Universal $Z^\prime$ model parameters.}
\label{fig:model-zprime}
\end{figure}

\begin{figure}[p]
\centering
\includegraphics[trim=0 0 0 20,width=.65\textwidth]{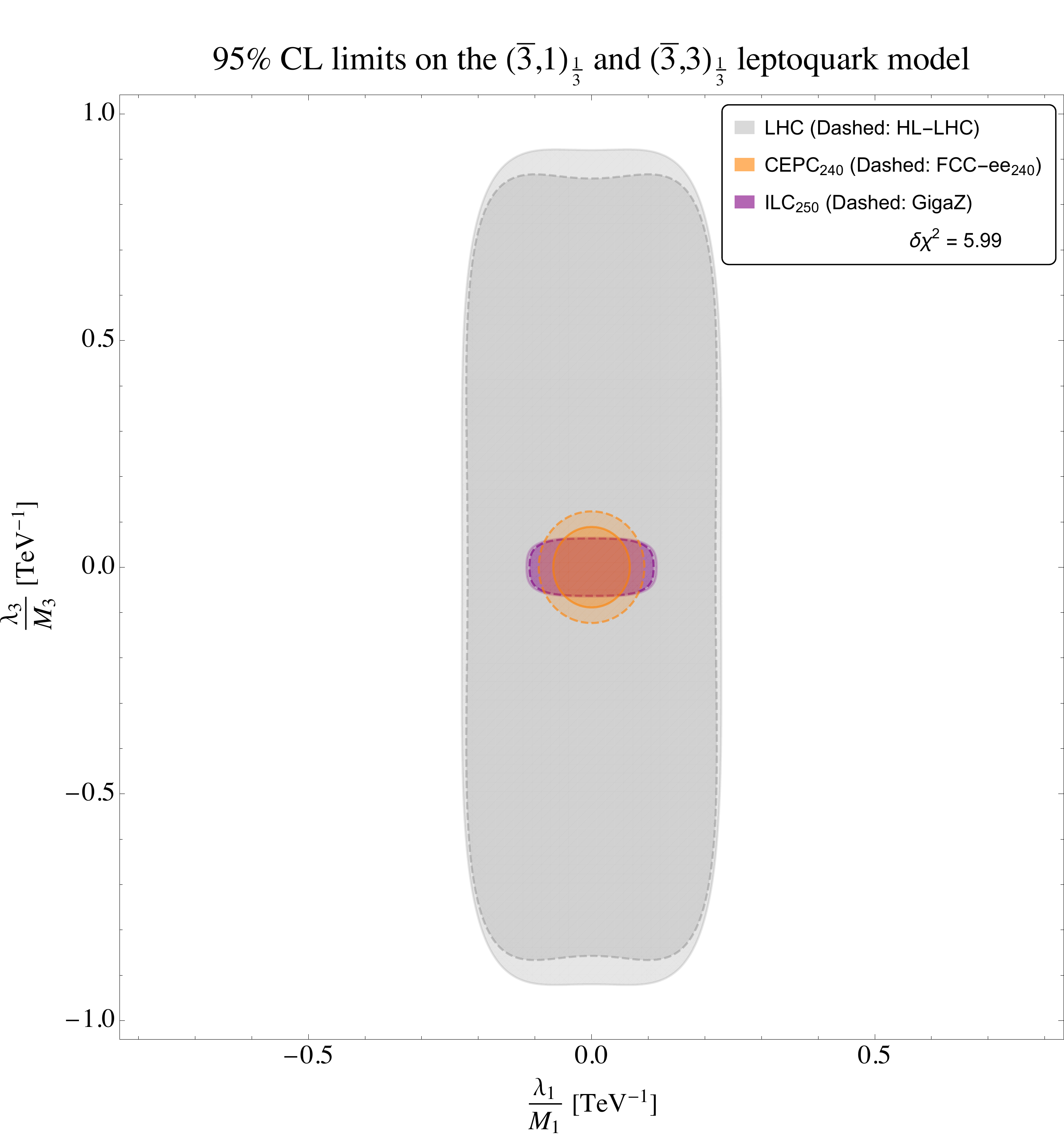}%
\vspace{-1ex}
\caption{95\% C.L. exclusion reach of different colliders on the leptoquark model parameters. Only future $e^+e^-$ scenarios with energies below the $t\bar{t}$ threshold have been considered since the analysis did not include any top-quark observables.}
\label{fig:model-leptoquark}
\end{figure}

%%%%%%%%%%%%%%%%%%%%%%%%%%%%%%%%%%%%%%%%%%%%%%%%%%%%%%%%%%%%%%%%%%%%%%%%%%%
%%%%%%%%%%%%%%%% Conclusions

\clearpage

\section{Conclusions}

\begin{itemize}
    \item For ``canonical'' electroweak precision measurements (Z-pole, WW threshold), circular $e^+e^-$ colliders (FCC-ee, CEPC) have in general a higher sensitivity than linear colliders (ILC, CLIC) due to the high luminosity at center-of-mass energies below 200~GeV. Beam polarization at the linear colliders improves their sensitivity and can help to control systematics. In particular, for a linear collider run on the Z pole, beam polarization would enable measurements of the asymmetry parameters $A_f$ with a precision that is only a factor of a few worse than for circular colliders, in spite of several orders of magnitude larger statistics for Z-pole physics at circular colliders.
    \item At center-of-mass energies $\sqrt{s} \lesssim 160$~GeV, the beam energy can be precisely calibrated using resonant depolarization at circular $e^+e^-$ colliders, thus enabling very precise determinations of Z and W masses and widths. Linear colliders need a physical mass for energy calibration, which could be the Z mass (with 25 ppm precision from LEP) or possibly hadron (kaon and $\Lambda$) masses. Using the latter may put an energy calibration with 2ppm precision within reach, but requires further investigation. 
    \item For many of the most precisely measurable precision observables at linear colliders, the most significant source of experimental systematics stems from the polarization calibration. For the circular colliders, on the other hand, modeling uncertainties for hadronic final states appear to be the dominant systematic error source.
    \item All $e^+e^-$ Higgs factory colliders are similarly affected by a class of systematic uncertainties due to QCD and hadronization modeling, in particular for heavy-flavor final states.
    \item At any proposed $e^+e^-$ collider it will be possible to measure the W mass with a precision of a few MeV or even better, thus conclusively resolving the recent discrepancy among different W mass determinations at hadron colliders \cite{CDF:2022hxs}.
    \item Experiments at lower-energy $e^+e^-$ colliders, lepton-proton or lepton-ion colliders, or neutrino scattering facilities can deliver complementary information about electroweak quantities, such as the running electroweak mixing angle at low scales, or the separate determination of up- and down-quark electroweak couplings.
    \item High-energy lepton colliders ($e^+e^-$ or $\mu^+\mu^-$ with $\sqrt{s} > 1\,\TeV$) are effectively boson colliders. The total cross-section for many production processes is dominated by VBF/VBS-type contributions. However, for studies of BSM effects at very high invariant masses, non-VBF processes become typically more dominant.
    \item At multi-TeV lepton colliders, multiple electroweak gauge-boson production is ubiquitous, and new theoretical tools are needed for calculating and simulating these effects.
    \item Hadron colliders and lepton colliders offer complementary information about potential new BSM physics: Measurements of EWPOs at future Higgs factories offer indirect sensitivity to heavy new physics at scales of several TeV, which in many cases substantially exceeds the reach of LHC / HL-LHC. Furthermore, they have unique sensitivity to very weakly coupled new particles with smaller masses. A hadron collider with 100~TeV center-of-mass energy, on the other hand, is able to directly produce new particles in the parameter space covered by indirect EWPO tests for most BSM scenarios (see EF BSM report \cite{Bose:2022obr} for more information). However, multi-TeV $e^+e^-$ or $\mu^+\mu^-$ colliders, while having lower statistical power than a $\O(100)$-TeV $pp$ collider, can have an advantage for certain multi-boson studies due to the well-defined initial state and clean event signatures. The specific benefits of hadron and lepton colliders depend on the type of BSM physics, and thus a combination of both collider types is needed for the broadest coverage of new physics scenarios.
    \item Assuming that any new particles are heavy, a model-independent parametrization of the new-physics reach of different colliders is given by the SMEFT framework, where the SM is extended by higher-dimensional operators, with the leading contribution for most processes entering at dimension 6. Several subsets of such dimension-6 operators have been investigated in a set of global fits across a large number of observables: (a) operators contributing to electroweak gauge-boson interactions; (b) operators contributing to Higgs interactions; (c) operators contrbuting to top-quark interactions; and (d) operators contributing to four-fermion contact interactions.
    \item Generally, future lepton colliders have a better reach for many of the aforementioned operators than the HL-LHC. Circular $e^+e^-$ colliders have the best sensitivity to electroweak operators, due to the large statistical precision of Z pole and WW threshold measurements. All lepton colliders are comparable in their reach for Higgs operators, although a multi-TeV muon collider cannot constrain exotic Higgs decays in a model-independent way\footnote{At future $e^+e^-$ Higgs factories or a 125-GeV muon collider run, a model-independent study of exotic Higgs decays is possible.}. Top-quark and four-fermion operators are best constrained at machines with $\sqrt{s} \geq 500$~GeV, and measurements at two or more values of $\sqrt{s}$ are crucial for breaking degeneracies. Many constraints on top-quark operators are improved by combining $e^+e^-$ and (HL-)LHC inputs and exploiting synergies between them.
    \item Some of the same SMEFT operators contribute to EW precision quantities and to Higgs observables or anomalous gauge-boson coupings (aGCs). At circular $e^+e^-$ colliders, measurements from a high-luminosity Z-pole run can improve the constraints on Higgs couplings and aGCs by a factor of up to 2. At linear $e^+e^-$ colliders, beam polarization provides additional information for Higgs/aGC measurements, and inputs from Z-pole data are less important. 
    \item Low-energy measurements (below the W/Z mass scales) are needed to close the fit for four-fermion operators, when allowing non-universality among the three fermion generations.
    \item At this point, not enough information was available to include $pp$ colliders beyond the LHC (such as HE-LHC or a $\O(100)$-TeV collider) in the global fit. It is likely that these machines have superior sensitivity to many energy-dependent operators, such as 4-fermion operators involving quarks and several operators that mediate multi-boson interactions.
\end{itemize}

%%%%%%%%%%%%%%%% Acknowledgments

\section*{Acknowledgments}

The conveners are grateful to C.~Grojean and M.~Shapiro for detailed reviews of the draft report, and to S.~Eno for many helpful comments.

%%%%%%%%%%%%%%%% References

\bibliographystyle{JHEP}
\bibliography{bibliography}

\end{document}